\documentclass[aps,prl,twocolumn,footinbib]{revtex4-1}
\usepackage{graphicx}
\usepackage{indentfirst}
\usepackage{braket}
\usepackage{float}
\usepackage{amsmath}
\usepackage{epstopdf}
\usepackage{CJK}
\usepackage{esint}
\usepackage{color}
\usepackage[T1]{fontenc}
\usepackage{subfigure}
\usepackage{letltxmacro}
\LetLtxMacro{\ORIGselectlanguage}{\selectlanguage}
\makeatletter
\DeclareRobustCommand{\selectlanguage}[1]{%
  \@ifundefined{alias@\string#1}
    {\ORIGselectlanguage{#1}}
    {\begingroup\edef\x{\endgroup
       \noexpand\ORIGselectlanguage{\@nameuse{alias@#1}}}\x}%
}
\newcommand{\definelanguagealias}[2]{%
  \@namedef{alias@#1}{#2}%
}
\makeatother

\definelanguagealias{en}{english}
\usepackage{amsfonts}
\usepackage{footmisc}
\usepackage{scrextend}
\usepackage{multirow}
\usepackage{cleveref}
\usepackage[english]{babel}
\usepackage{url}

\newcommand{\qed}{\nobreak \ifvmode \relax \else
      \ifdim\lastskip<1.5em \hskip-\lastskip
      \hskip1.5em plus0em minus0.5em \fi \nobreak
      \vrule height0.75em width0.5em depth0.25em\fi}

\def\be{\begin{equation}}
\def\ee{\end{equation}}
\def\ba{\begin{eqnarray}}
\def\ea{\end{eqnarray}}

\begin{document}
\begin{CJK*}{UTF8}{gbsn}
\title{Optimum mixed-state discrimination for noisy entanglement-enhanced sensing}

\author{Quntao Zhuang$^{1,2}$}
\email{quntao@mit.edu}
\author{Zheshen Zhang$^1$}
\author{Jeffrey H. Shapiro$^1$}
\affiliation{$^1$Research Laboratory of Electronics, Massachusetts Institute of Technology, Cambridge, Massachusetts 02139, USA\\
$^2$Department of Physics, Massachusetts Institute of Technology, Cambridge, Massachusetts 02139, USA}
\date{\today}
\begin{abstract}

Quantum metrology utilizes nonclassical resources, such as entanglement or squeezed light, to realize sensors whose performance exceeds that afforded by classical-state systems. Environmental loss and noise, however, easily destroy nonclassical resources, and thus nullify the performance advantages of most quantum-enhanced sensors. Quantum illumination (QI) is different.  It is a robust entanglement-enhanced sensing scheme whose 6\,dB performance advantage over a coherent-state sensor of the same average transmitted photon number survives the initial entanglement's eradication by loss and noise. Unfortunately, an implementation of the optimum quantum receiver that would reap QI's full performance advantage has remained elusive, owing to its having to deal with a huge number of very noisy optical modes. We show how sum-frequency generation (SFG) can be fruitfully applied to optimum multi-mode Gaussian-mixed-state discrimination.   Applied to QI, our analysis and numerical evaluations demonstrate that our SFG receiver saturates QI's quantum Chernoff bound.  Moreover, augmenting our SFG receiver with a feed-forward (FF) mechanism pushes its performance to the Helstrom bound in the limit of low signal brightness. The FF-SFG receiver thus opens the door to optimum quantum-enhanced imaging, radar detection, state and channel tomography, and communication in practical Gaussian-state situations. 

\end{abstract}

\maketitle
\end{CJK*}

{\em Introduction.---} Entanglement is essential for device-independent quantum cryptography \cite{Ekert_1991}, quantum computing \cite{Shor_1997}, and quantum-enhanced metrology \cite{Wineland_1992}. It has also been employed in frequency and phase estimation to beat their standard quantum limits on measurement precision \cite{Dowling_1998,Afek_2010,Xiang_2011,Ono_2013,Anisimov_2010,Yonezawa_2012,Czekaj_2015}. 
Furthermore, entanglement has applications across diverse research areas, including dynamic biological measurement \cite{Biological}, delicate material probing \cite{Wolfgramm_delicate_material}, gravitational wave detection \cite{LIGO_2}, and quantum lithography \cite{Lithography}.
Entanglement, however, is fragile; it is easily destroyed by quantum decoherence arising from environmental loss and noise. Consequently, the entanglement-enabled performance advantages of most quantum-enhanced sensing schemes quickly dissipate with increasing quantum decoherence, challenging their merits for practical situations.

Quantum illumination (QI) is an entanglement-enhanced paradigm for target detection that thrives on entanglement-breaking loss and noise \cite{Sacchi_2005_1,Sacchi_2005_2,Lloyd2008,Tan2008,Lopaeva_2013,Guha2009,Zheshen_15,Barzanjeh_2015}. Its optimum quantum receiver enjoys a 6\,dB advantage in error-probability exponent over optimum classical sensing using the same transmitted photon number.  Remarkably, QI's advantage occurs despite the initial entanglement being completely destroyed. 

To date, the only in-principle realization of QI's optimum quantum receiver requires a Schur transform on a quantum computer \cite{Bacon_2006}, so that its physical implementation is unlikely to occur in the near future.  At present, the best known sub-optimum QI receivers \cite{Guha2009,Zheshen_15}---one of which, the optical parametric amplifier (OPA) receiver, has been demonstrated experimentally \cite{Zheshen_15}---can only realize a 3\,dB error-probability exponent advantage. Bridging the 3\,dB performance gap between the sub-optimum and optimum receivers with an implementation more feasible than a quantum computer is of particular significance for its application potential and for its deepening our understanding of entanglement-enhanced metrology.

In this Letter we present an optimum QI-receiver architecture based on sum-frequency generation (SFG). In the weak-signal limit, the SFG unitary maps QI target detection to the well-studied problem of single-mode coherent state discrimination (see Ref.~\cite{SuppMaterial} for a review). Analytical calculation and Monte Carlo simulations confirm that this SFG receiver's performance approaches QI's quantum Chernoff bound (QCB) \cite{Tan2008} asymptotically.  Adding a feed-forward (FF) mechanism yields the FF-SFG receiver, whose error probability achieves the Helstrom bound \cite{Helstrom1969}.  The FF-SFG receiver is potentially promising for other quantum-enhanced sensing scenarios, such as phase estimation, and it enlarges the toolbox for quantum-state discrimination \cite{Chefles_2000,Barnett_09,Barnett_1998,Andersson_2002,Anthony_1998,Chen_2015,BecerraF_2013,Becerra_2015,Mosley_2006,Clarke_2001,Tsujino_2011,Takeoka_2006,Takeoka_2005,Nair_2014}. In particular, it is the first architecture---short of a quantum computer---for optimum discrimination of multi-mode Gaussian mixed states, a major step beyond the optimum discrimination of single-mode pure states \cite{Kennedy_1972,dolinar_processing_1973,Sasaki_Hirota_receiver,Acin_2005}.

{\em Target detection.---} QI target detection works as follows \cite{Tan2008}.   An entanglement source generates $M \gg 1$ signal-idler mode pairs, having photon annihilation operators $\{\hat{c}_{S_{0_m}},\hat{c}_{I_{0_m}} : 1\le m\le M\}$, with each pair being in a two-mode squeezed-vacuum state of mean photon number $2N_S \ll 1$. The signal modes probe for the presence of a weakly-reflecting target embedded in a bright background, under the assumption that it is equally likely to be absent or present, while the idler modes are retained for subsequent joint measurement with light collected from the region interrogated by the signal modes.  (We shall assume lossless idler storage, so that the idler modes used for that joint measurement satisfy $\hat{c}_{I_m} = \hat{c}_{I_{0_m}}$.)  When the target is present (hypothesis $h = 1$), the returned signal modes are $\hat{c}_{S_m} = \sqrt{\kappa}\,\hat{c}_{S_{0_m}}+\sqrt{1-\kappa}\,\hat{c}_{N_m}$, where $\kappa\ll1$ is the roundtrip transmissivity and the $\{\hat{c}_{N_m}\}$ are noise modes in thermal states containing $N_B /(1-\kappa) \gg 1$ photons on average. When the target is absent (hypothesis $h = 0$), the returned signal modes are $\hat{c}_{S_m} = \hat{c}_{N_m}$, where the $\{\hat{c}_{N_m}\}$ are now taken to be in thermal states with average photon number $N_B$ \cite{footnote0}.  

Omitting the $\kappa N_S\ll N_B$ contribution to $\langle \hat{c}_{S_m}^\dagger\hat{c}_{S_m}\rangle$ when the target is present, and conditioned on $h=j$, the $\{\hat{c}_{S_m},\hat{c}_{I_m}\}$ constitute a set of independent, identically-distributed (iid) mode pairs that are in zero-mean Gaussian states with Wigner covariance matrix 
\begin{align}
& 
{\mathbf{\Lambda}}_j =
\frac{1}{4}
\left(
\begin{array}{cccc}
(2N_B+1) {\mathbf I}&2C_p{\mathbf Z}\delta_{1j}\\
2C_p {\mathbf Z}\delta_{1j}&(2N_S+1){\mathbf I}
\end{array} 
\right),
&
\label{CovReturnIdler_main}
\end{align}
where ${\mathbf I} = {\rm diag}(1,1)$, ${\mathbf Z}={\rm diag}(1,-1)$, $\delta_{ij}$ is the Kronecker delta function, and $C_p = \sqrt{\kappa N_S(N_S+1)}$ is the phase-sensitive cross correlation that exists when the target is present.   The task of QI target detection is thus minimum error-probability discrimination between two $M$-mode-pair, zero-mean Gaussian states that are characterized by the $\{{\mathbf{\Lambda}}_j\}$. 

For equally-likely hypotheses, the minimum error-probability quantum measurement for discriminating between states with density operators $\hat{\rho}_0$ and $\hat{\rho}_1$ is the Helstrom measurement $u(\hat{\rho}_1-\hat{\rho}_0)$, where $u(x) = 1$ for $x\ge 0$ and 0 otherwise  \cite{Helstrom1969}.  Absent the availability of a quantum computer, the best known QI receivers have error-probability exponents that are 3\,dB inferior to optimum quantum reception. These sub-optimum receivers use Gaussian local operations on each mode pair plus photon-number resolving measurements, and hence belong to the class of local operations plus classical communication (LOCC). Their sub-optimality follows because LOCC is not optimum for general mixed-state discrimination \cite{LOCC_NO_GO,Bandyopadhyay_2011}.  

To go beyond LOCC, we will employ SFG.  The QI transmitter uses a continuous-wave spontaneous parametric downconverter (cw-SPDC) to generate $M \gg 1$ signal-idler mode pairs---at frequencies $\{\omega_{S_m},\omega_{I_m}\}$---during target-region interrogation. These mode pairs originate from a single-mode pump $\hat{b}$ at frequency $\omega_b= \omega_{S_m}+\omega_{I_m}$.  Each mode has average photon number $N_S$ and each mode pair has a phase-sensitive cross correlation $\sqrt{N_S(N_S+1)}$.  SFG is SPDC's inverse process:  $M$ independent signal-idler mode pairs with the same phase-sensitive cross correlation can combine, coherently, to produce photons at the pump frequency.  It is natural, therefore, to explore SFG in seeking an optimum QI receiver, because the phase-sensitive cross correlation $C_p$ in Eq.~(\ref{CovReturnIdler_main}) is the signature of target presence.  We begin with some foundational results for SFG.  

{\em Sum-frequency generation.---}
We will describe SFG by Schr\"{o}dinger evolution for $t\ge 0$ under interaction Hamiltonian
\be
\hat{H}_I = \hbar g\sum_{m=1}^{M} (\hat{b}^\dagger \hat{a}_{S_m}\hat{a}_{I_m} +\hat{b}\hat{a}_{S_m}^\dagger\hat{a}_{I_m}^\dagger ),
\label{Hamiltonian_main}
\ee
with $M\gg 1$, where $\hbar$ is the reduced Planck constant and $g$ is the interaction strength.  We will assume that at time $t=0$ the $\{\hat{a}_{S_m},\hat{a}_{I_m}\}$ mode pairs (at frequencies $\{\omega_{S_m},\omega_{I_m}\}$) are in iid zero-mean Gaussian states, while the $\hat{b}$ sum-frequency mode (at frequency $\omega_b = \omega_{S_m}+\omega_{I_m}$) is in its vacuum state.  We will assume that the state evolution stays wholly within the low-brightness, weak cross-correlation regime wherein $n_s(t) \equiv \langle\hat{a}^\dag_{S_m}\hat{a}_{S_m} \rangle_t \ll 1$, $n_i(t) \equiv \langle\hat{a}^\dag_{I_m} \hat{a}_{I_m}\rangle_t\ll 1$, and $|C(t)|^2 \equiv |\braket{\hat{a}_{S_m}\hat{a}_{I_m}}_t|^2 \ll n_s(t),n_i(t)$ for all time, where $\langle \cdot\rangle_t$ denotes averaging with respect to the state at time $t$.  The qubit approximation to this evolution leads to the analytical results \cite{SuppMaterial}
\begin{subequations}
\label{SFG_evolution_main}
\ba
C(t)&=&C(0)\cos (\sqrt{M}gt)
\\
b(t)&=&-i \sqrt{M} C(0) \sin ( \sqrt{M}gt)
\\
n_s(t)&=&n_s(0),
n_i(t)=n_i(0)
\\
n_b(t)&=&\left[M|C(0)|^2+n_i(0)n_s(0)\right]\sin^2(\sqrt{M}gt),
\label{n_b_evolution_main}
\ea
\end{subequations}
where $b(t)\equiv\braket{\hat{b}}_t$ and $n_b(t)\equiv \langle\hat{b}^\dag\hat{b}\rangle_t$. The average photon numbers in the $\{\hat{a}_{S_m},\hat{a}_{I_m}\}$ are constant, in this approximation, because each mode's $n_b(t)/M$ contribution to the sum-frequency mode's average photon number is negligible. Equations~(\ref{SFG_evolution_main}) agree very well with numerical results for $M=1,2,$ and $3$ \cite{SuppMaterial}. For any $M$ they reveal a coherent oscillation between the $\hat{b}$ mode's mean field and the cross correlation in {\em all} signal-idler mode pairs, plus an additional $M$-independent oscillation in the $\hat{b}$ mode's average photon number from the weak thermal-noise contribution ($\propto n_i(0)n_s(0)$), to $n_b(t)$. 

{\em Optimum receiver design.---} Were $\langle \hat{c}_{S_m}^\dagger\hat{c}_{S_m}\rangle \ll 1$ under both hypotheses, QI's returned-signal and retained-idler mode pairs would satisfy the low-brightness conditions needed for Eqs.~(\ref{SFG_evolution_main}) to apply.   Then, when these mode pairs undergo SFG with the sum-frequency mode $\hat{b}$ initially in its vacuum state, $\hat{b}$'s output state at $t=\pi/2\sqrt{M}g$ would be approximately a weak thermal state (average photon number $n_T = \langle \hat{c}_{I_m}^\dagger\hat{c}_{I_m}\rangle\langle \hat{c}_{S_m}^\dagger\hat{c}_{S_m}\rangle$) when $h=0$, or a coherent state (with mean field $-i\sqrt{M}C_p$) embedded in a weak thermal background (average photon number $n_T$) when $h=1$.  Minimum error-probability discrimination between the two hypotheses, based on $\hat{b}$'s output state, is then a single-mode Gaussian mixed-state problem \cite{SuppMaterial}.  Unfortunately, Eq.~(\ref{CovReturnIdler_main}) implies that $\langle \hat{c}_{S_m}^\dagger\hat{c}_{S_m}\rangle_0 = N_B \gg 1$ under both hypotheses, violating the low-brightness condition.  When these bright signal modes undergo SFG, they drive $\hat{b}$ to an equilibrium state \cite{Tanas91}, precluding the desired coherent conversion. 

\begin{figure}[htb]
\centering
\includegraphics[width=0.47\textwidth]{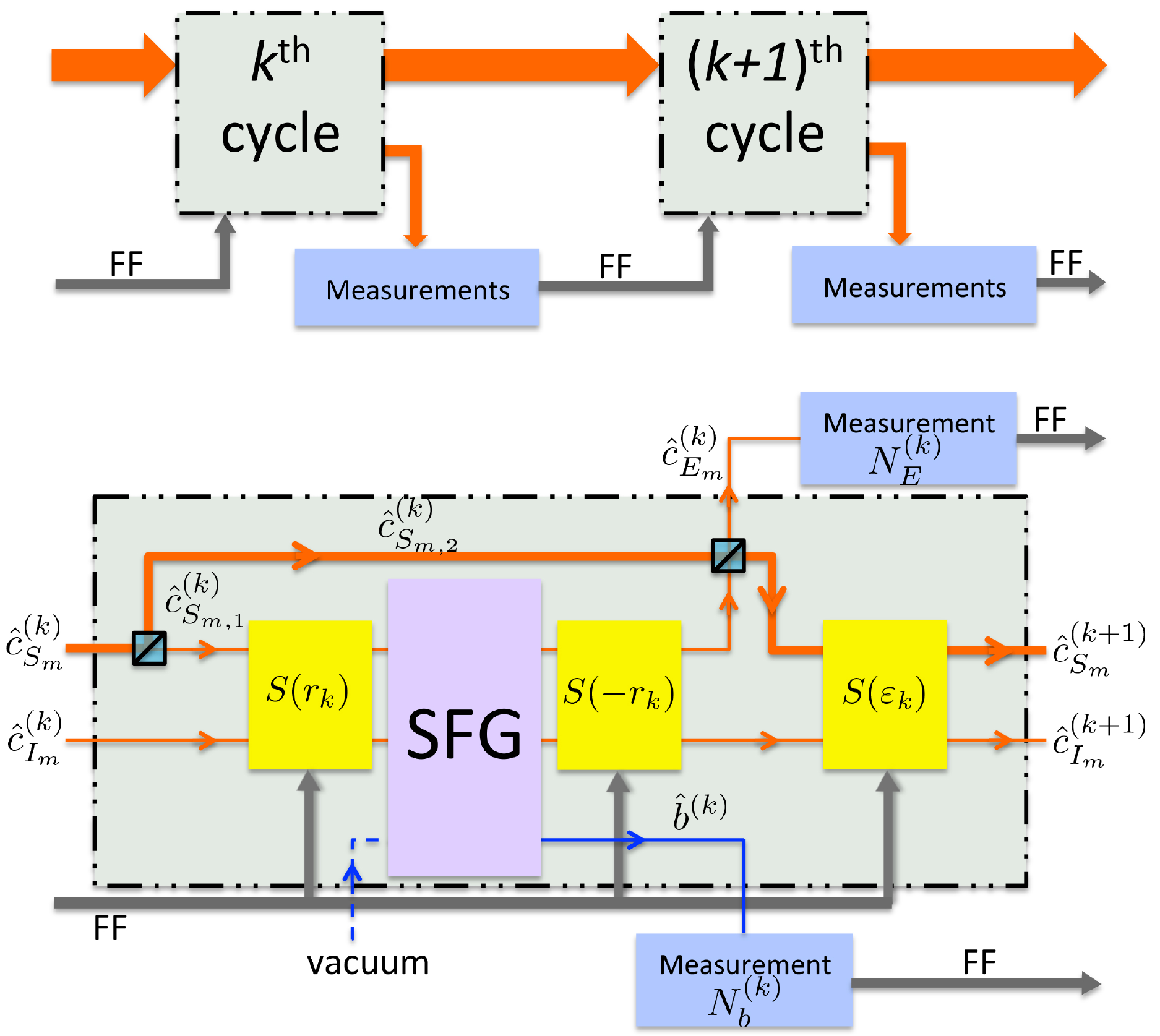}
\caption{Schematic of the FF-SFG receiver. Upper panel: two successive cycles. Lower panel: the components in the $k{\rm th}$ cycle. $S(\cdot)$: two-mode squeezing; SFG: sum-frequency generation; FF: feed-forward operation.
}
\label{Fig1main}
\end{figure}

To resolve this $N_B\gg 1$ problem, we propose a receiver that uses $K$ cycles of $\pi/2\sqrt{M}g$-duration SFG interactions, as shown in Fig.~\ref{Fig1main}.  With optimum choices of the $\{r_k, \varepsilon_k\}$, this figure represents the FF-SFG receiver; setting all the $\{r_k,\varepsilon_k\}$ to zero reduces it to the SFG receiver.   We shall describe the FF-SFG receiver, but present performance results for both receivers.  It suffices to consider a single cycle comprised of one SFG interaction, plus the pre-SFG signal slicing, the post-SFG signal combining, and the post-SFG photon-counting measurements.  

Let $\{\hat{c}_{S_m}^{(k)},\hat{c}_{I_m}^{(k)}\}$ be the signal-idler mode pairs at the input to the $k$th cycle for $0\le k \le K-1$, with $\hat{c}_{S_m}^{(0)} = \hat{c}_{S_m}$ and $\hat{c}_{I_m}^{(0)} = \hat{c}_{I_m}$.  A transmissivity $\eta\ll 1$ beam splitter taps a small portion of each $\hat{c}_{S_m}^{(k)}$ mode, yielding a low-brightness transmitted mode $\hat{c}_{S_m,1}^{(k)}$ to undergo a two-mode squeezing (TMS) operation $S(r_k)$ \cite{footnote}, with the $\hat{c}_{I_m}^{(k)}$ mode, and a high-brightness reflected mode $\hat{c}_{S_m,2}^{(k)}$ to be retained.  For the FF-SFG receiver, the $r_k$ value (which depends on $\tilde{h}_k = 0$ or 1, the minimum error-probability decision as to target absence or presence based on the measurement results from all prior cycles \cite{footnoteDecision}) is chosen to almost purge any phase-sensitive cross correlation between the $\{\hat{c}_{S_m,1}^{(k)},\hat{c}_{I_m}^{(k)}\}$ mode pairs from the $S(r_k)$ operation's output mode pairs were $\tilde{h}_k$ a correct decision.   Because $S(r_k)$'s output mode pairs are applied to an SFG process that converts any mode-pair phase-sensitive cross correlation to a non-zero mean field for its sum-frequency ($\hat{b}^{(k)}$) mode's output, any significant mean field indicates that the $\tilde{h}_k$ decision was wrong.  As shown in \cite{SuppMaterial}:  (1) $\hat{b}^{(k)}$ is not entangled with any other SFG output mode; and (2) each signal-idler mode pair emerging from SFG is in a Gaussian state. These facts allow us to use the weak TMS operation $S(\sqrt{\eta}C_{si}^{(k)}-r_k)$ to approximate the SFG operation on each signal-idler mode pair, where  $C_{si}^{(k)}\equiv\braket{\hat{c}_{S_m}^{(k)}\hat{c}_{I_m}^{(k)}}$.

Following the $k$th cycle's SFG operation, we apply the TMS operation $S(-r_k)$ to each signal-idler mode pair. Under either hypothesis, the number of photons lost by the signal modes entering the SFG operation matches the number of photons gained by the $\hat{b}^{(k)}$ mode. The $S(-r_k)$ operation ensures that, when its signal-mode outputs are combined with the  retained $\{\hat{c}_{S_m,2}^{(k)}\}$ modes on a second transmissivity-$\eta$ beam splitter, the $\{\hat{c}_{E_m}^{(k)}\}$ output modes contain the same number of photons as the $\hat{b}^{(k)}$ mode.  The photon-number measurements $\hat{b}^{(k)\dagger}\hat{b}^{(k)}$ and $\sum_{m=1}^M\hat{c}_{E_m}^{(k)\dagger}\hat{c}_{E_m}^{(k)}$ then provide outcomes $N_b^{(k)}$ and $N_E^{(k)}$ that are substantial when $\tilde{h}_k$ is incorrect, but negligible when $\tilde{h}_k$ is correct. These measurement outcomes are fed-forward for use in determining $\tilde{h}_{k+1}$, with  $\tilde{h}_K$ being the receiver's final decision as to whether the target is absent or present.

The $k$th cycle is completed by a TMS operation $S(\varepsilon_k)$, with $\varepsilon_k=\sqrt{\eta}\,r_k$,  that makes the phase-sensitive cross correlation of the signal and idler inputs to the $(k+1)$th cycle independent of $r_k$.   The first-order results for the conditional moments given $h=j$ are \cite{SuppMaterial}:
\begin{subequations}
\label{SFGcycle_main}
\ba
n_s^{(k)}&\equiv& \langle \hat{c}_{S_m}^{(k)\dagger}\hat{c}_{S_m}^{(k)}\rangle|_{h=j} = N_B\\
n_i^{(k)}&\equiv& \langle \hat{c}_{I_m}^{(k)\dagger}\hat{c}_{I_m}^{(k)}\rangle|_{h=j} =  N_S
\\
C_{si}^{(k)}|_{h=j} &=& j C_p[1-\eta(1+N_B)]^k.
\label{Csi_solution_main}
\ea
\end{subequations}

{\em Feed-forward and decision.---} All that remains to fully specify the FF-SFG receiver is to derive the optimum $\{r_k\}$ and $\{\tilde{h}_k\}$ values, and to choose an appropriate value for $K$, the number of cycles to be employed. To do so, we will draw on a connection to Dolinar's optimum receiver for binary coherent-state discrimination \cite{dolinar_processing_1973}
by setting $r_k = 0$, to consider the SFG receiver, and omitting the small incoherent contribution to the $\hat{b}^{(k)\dagger}\hat{b}^{(k)}$ measurement. Then, assuming $h=1$, the $k$th cycle produces a $\hat{b}^{(k)}$ mode in a coherent state with average photon number $\langle N_b^{(k)}\rangle |_{h=1}=M\lambda_k^2$ and $\{\hat{c}_{E_m}^{(k)}\}$ modes in iid thermal states with total average photon number $\langle N_E^{(k)}\rangle |_{h=1}=M\lambda_k^2$, where $\lambda_k\equiv\sqrt{\eta}\,C_{si}^{(k)}|_{h=1}$. For $\eta$ sufficiently small, the $h=1$ statistics of $N^{(k)}\equiv N_b^{(k)} + N_E^{(k)}$ will match the photon-number statistics of the coherent state $|\sqrt{2M}\lambda_k\rangle$. On the other hand, the $h=0$ statistics of $N^{(k)}$ are those of the vacuum state, i.e., $N^{(k)} = 0$ with probability one. Optimum binary coherent-state discrimination \cite{dolinar_processing_1973,Acin_2005} applied to our problem then gives $r_{k} = r^{(k)}_{\tilde{h}_k}$, where (see Ref.~\cite{SuppMaterial} for an intuitive explanation)
\be
\label{rk}
r^{(k)}_{\tilde{h}_k} =  \frac{\lambda_k}{2}\!\left(1-\frac{(-1)^{\tilde{h}_k}}{\sqrt{1-\exp\left[-2 M(\sum_{\ell=0}^{k}\lambda_\ell^2-\lambda_k^2/2)\right]}}\right).
\ee
Here, $\tilde{h}_k$ is the $j$ value that maximizes $P^{(k)}_{h=j}$ \cite{footnoteDecision}, where the prior probabilities for the $k$th cycle, $\{P^{(k)}_{h=j} : j = 0, 1\}$, are the posterior probabilities of the $\left(k-1\right)$th cycle that are obtained from the Bayesian update rule~\cite{Acin_2005,Assalini2011}, 
\be
P_{h=j}^{(k)}=\frac{P_{h=j}^{(k-1)}P_{BE}(N_b^{(k-1)},N_E^{(k-1)};j,r_{\tilde{h}_{k-1}}^{(k-1)})}{\sum_{j=0}^1 P_{h=j}^{(k-1)}P_{BE}(N_b^{(k-1)},N_E^{(k-1)};j,r_{\tilde{h}_{k-1}}^{(k-1)})},
\ee
for $1\le k \le K-1$, where $P_{BE}(N_b^{(k-1)},N_E^{(k-1)};j,r_{\tilde{h}_{k-1}}^{(k-1)})$ is the conditional joint probability of getting counts $N_b^{(k-1)}$ and $N_E^{(k-1)}$ given that the true hypothesis is $j$ and $r_{k-1} = r_{\tilde{h}_{k-1}}^{(k-1)}$. The $S(r_{k-1})$-SFG-$S(-r_{k-1})$ cascade in the $(k-1)$th cycle is designed to make the photon fluxes that generate $N_b^{(k-1)}$ and $N_E^{(k-1)}$ much higher if $\tilde{h}_{k-1} \neq h$ than if $\tilde{h}_{k-1} = h$. Thus the update rule will flip $\tilde{h}_k$ to the other hypothesis if too many photons are counted in the $(k-1)$th cycle; otherwise $\tilde{h}_k = \tilde{h}_{k-1}$ will prevail.    

To determine how many cycles must be run, we reason as follows.  Suppose that $h=1$ and we continue to neglect the small incoherent contribution to the $\hat{b}^{(k)\dagger}\hat{b}^{(k)}$.  We then have that $N_T^{(K)} \equiv \sum_{k=0}^{K-1}N^{(k)} = 2M\sum_{k = 0}^{K-1}\lambda_k^2$ is the total average photon number of \emph{all} the measurements made in the FF-SFG receiver's $K$ cycles.  To ensure that the receiver's final decision, $\tilde{h}_K$, as to whether the target is absent ($\tilde{h}_K = 0$) or present ($\tilde{h}_K = 1$) is optimum, two conditions should be satisfied:  (1) $\eta$ is small enough that the qubit approximations in \cite{SuppMaterial} are valid; and (2) $K$ is large enough that $N_T^{(K)}/N_T^{(\infty)} = 1-\epsilon$, for some pre-chosen $0< \epsilon \ll 1$.  

{\em Performance.---}
We begin our performance evaluations for the FF-SFG and SFG receivers with some asymptotic results \cite{SuppMaterial}.  For $\eta$ sufficiently small, the coherent and incoherent (thermal-state) contributions to $N_T^{(K)}$ are $N_{T_{\rm coh}}^{(K)}\simeq (1-\epsilon)M\kappa N_S/N_B$ and $N_{T_{\rm therm}}^{(K)}\simeq -N_S\ln(\epsilon)/2$, and the number of cycles employed is 
$
K\simeq{-\ln(\epsilon)}/2\eta N_B
$.
Equations~(\ref{SFGcycle_main}), which underlie these expressions, are valid only when $N_S\ll1$. So, to get asymptotic results, we let $N_S\to 0$, to drive $N_{T_{\rm therm}}^{(K)}$ to zero, and we increase the source's mode number, $M$, to keep $N_{T_{\rm coh}}^{(K)}$ constant.   In this regime, QI target detection with the FF-SFG and SFG receivers becomes one of discriminating the coherent state $|\sqrt{N_{T_{\rm coh}}^{(K)}}\rangle$ from the vacuum.  Like the case for the Dolinar receiver \cite{dolinar_processing_1973}, the FF-SFG receiver's error probability should then approach the Helstrom bound 
$
P_H= \left[1-\sqrt{1-\exp(-N_{T_{\rm coh}}^{(K)})}\right]/2,
$
and, like the case for the Kennedy receiver \cite{Kennedy_1972}, the SFG receiver's error-probability exponent should approach $N_{T_{\rm coh}}^{(K)}$, which, for $\epsilon\to 0$, is both the QCB for the preceding coherent-state discrimination problem \emph{and} that for QI target detection. 

\begin{figure}[htb]
\centering
\subfigure{
\includegraphics[width=0.225\textwidth]{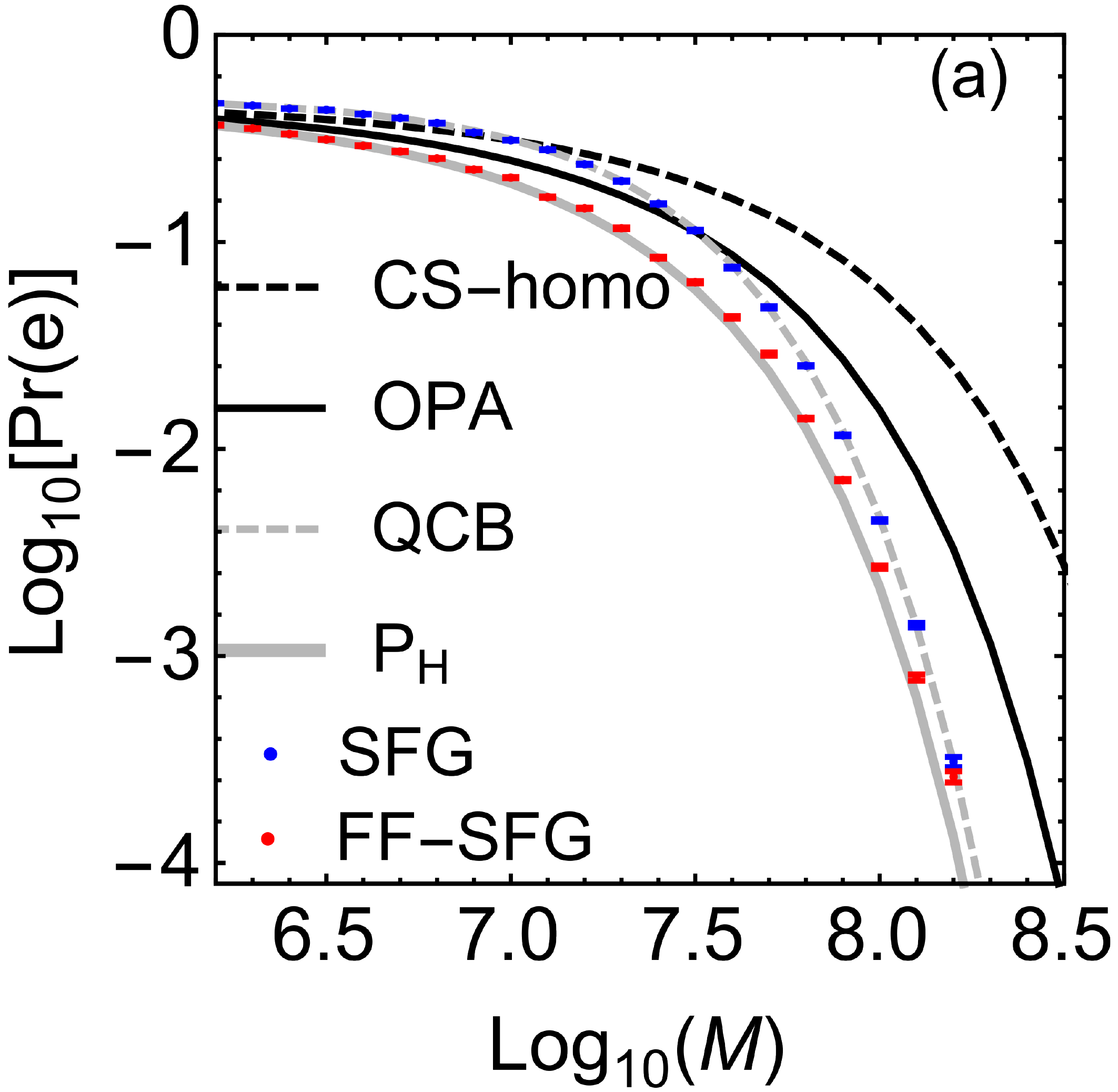}
\label{Fig2amain}
}
\subfigure{
\includegraphics[width=0.227\textwidth]{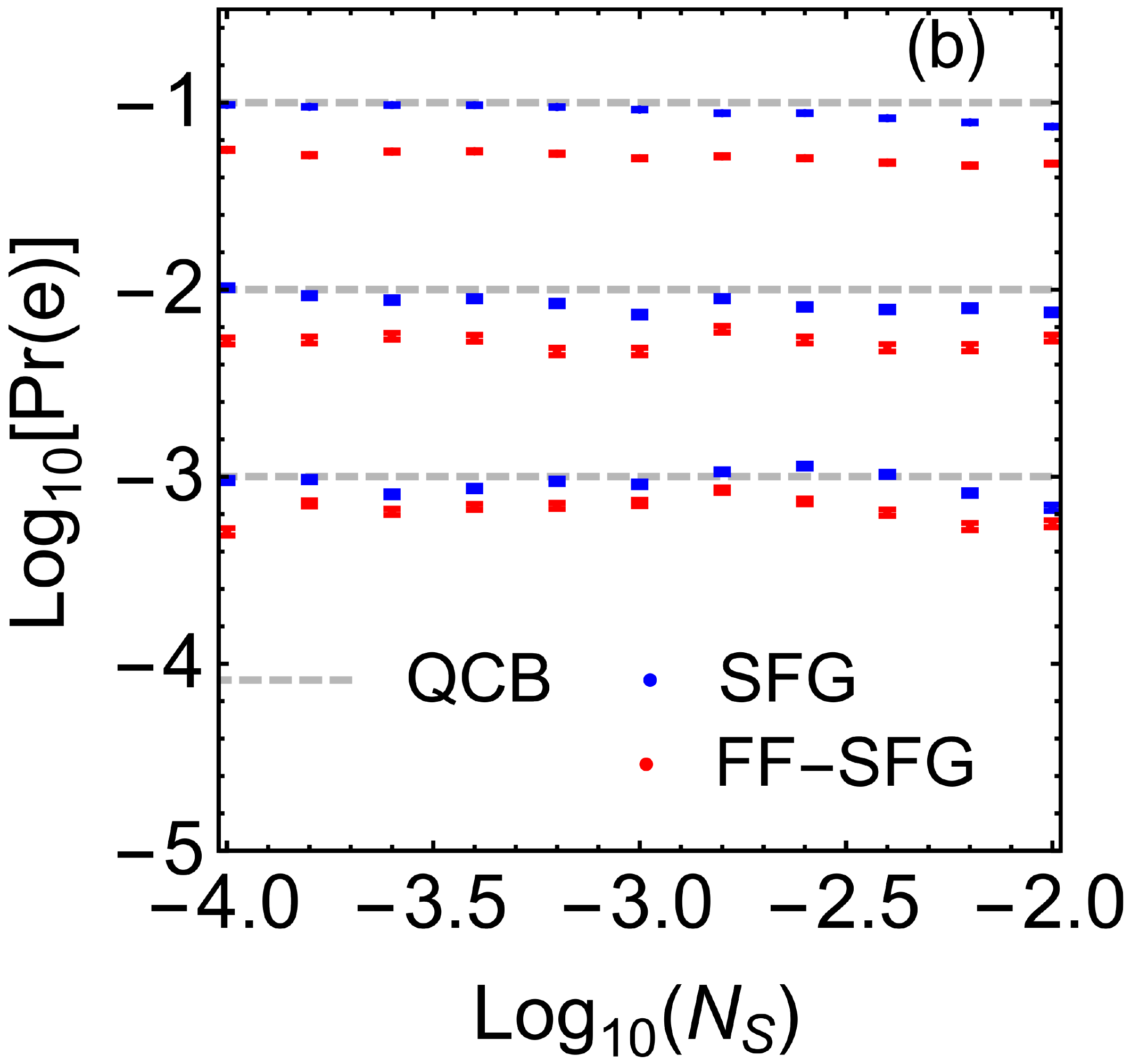}
\label{Fig2bmain}
}
\caption{(a) Error probabilities for the SFG, FF-SFG, and OPA receivers obtained from Monte Carlo simulations, plus analytical results for coherent-state discrimination with a homodyne receiver, and the Helstrom limit $P_H$ when $N_{T_{\rm coh}}^{(K)}=M\kappa N_S/N_B$.  Parameter values are given in the text.
(b) Error-probability exponents for the SFG and FF-SFG receivers versus source brightness, $N_S$, with $M$ is chosen to keep the QI target-detection QCB at (top to bottom) $10^{-1}, 10^{-2}$, or $10^{-3}$.  Simulations run were $10^6$ for QCB = $10^{-3}$ and $10^5$ otherwise.
}
\label{Fig2main}
\end{figure}

To explore how closely the FF-SFG and SFG receivers' error probabilities approach their asymptotic behavior we performed Monte Carlo simulations using $N_S = 10^{-4}$, $\kappa=0.01$, $N_B = 20$, $\eta = 0.002$,  and $K = 42$.  These parameter values are consistent with the qubit approximation's validity.  We  used $10^5$ (for $\log_{10}M<7.8$) to $10^6$ simulation runs (for $\log_{10}M\ge 7.8$) to obtain our error-probability estimates \cite{SuppMaterial}.   Figure~\ref{Fig2amain} compares $M$-dependent simulation results for the error probabilities of the FF-SFG, SFG, and OPA receivers with those of the homodyne receiver for coherent-state discrimination and the Helstrom bound with $N_{T_{\rm coh}}^{(K)}=M\kappa N_S/N_B$.  At all $M$ values shown, both proposed receivers outperform the OPA receiver, with FF-SFG reception's performance approaching $P_H$. More importantly, our receivers asymptotically saturate the QCB.  Figure~\ref{Fig2bmain} shows Monte Carlo results comparing the error-probability exponents of the SFG and FF-SFG receivers with QI target-detection's QCB as a function of source brightness with $M$ chosen to keep the QCB constant   at $10^{-1}, 10^{-2},$ or $10^{-3}$. Increasing $N_S$ increases $N_{T_{\rm therm}}^{(K)}$, so Fig.~\ref{Fig2bmain} shows our receivers approach QCB performance over a wide range of noise values.

{\em Discussion.---} We have presented a structure for achieving asymptotically-optimum performance in QI target detection.  Compared to the Schur-transform approach to optimum mixed-state discrimination, the components of our FF-SFG and SFG receivers, albeit challenging, have simpler realizations.  In particular, the required SFG can be implemented in an optical cavity or nonlinear waveguides \cite{Guerreiro_2013}, and its $K$ cycles can be combined on a photonic integrated circuit \cite{Najafi_2015,Mower_2015,Carolan_2015}.  Feed-forward operations have been successfully employed to obtain improved performance in the discrimination of coherent states \cite{Chen_2015,BecerraF_2013,Becerra_2015}, mixed states \cite{Higgins_2009}, and entangled states \cite{Lu_2010}.   Furthermore, our receivers have other potential applications, including optimum reception for the QI communication protocol \cite{Shapiro2009}, and quantum state and channel tomography \cite{Lvovsky_2009,Acin_2001}.  

Three final points deserve mention.  First, our receiver's slicing approach is analogous to that in \cite{daSilva2013}, where it was shown that slicing could be used to achieve the Holevo capacity for classical information transmission over a pure-loss channel.  Second, recent work \cite{Wilde2016} has shown that QI offers a great performance advantage in target detection in the Neyman-Pearson setting, when the miss probability, $\Pr(\tilde{h}_K \neq h\mid h = 1)$, is to be minimized subject to a constraint on the false-alarm probabilitiy, $\Pr(\tilde{h}_K \neq h\mid h= 0)$.  The optimum quantum measurement for Neyman-Pearson detection,  $u(\hat{\rho}_1-\zeta\hat{\rho}_0)$ for an appropriately chosen real-valued $\zeta$, is identical to that for minimum error-probability discrimination between $\hat{\rho}_1$ and $\hat{\rho}_0$ when $\zeta = \Pr(h=0)/\Pr(h=1)$.  Thus, just as the Dolinar receiver can be initialized to achieve the Helstrom bound for coherent-state discrimination with unequal priors and hence for Neyman-Pearson discrimination, so too can our FF-SFG receiver for QI target detection.  Finally, we note that the implementation burden on our FF-SFG receiver can be vastly reduced by replacing its feed-forward stages with feedback stages, i.e., we implement only one cycle and feed back its optical outputs to its inputs while using its measurement outputs to adjust its $r_k$ and $\varepsilon_k$ values.  Running this feedback arrangement through $K$ cycles then yields the same output as the original feed-forward setup but with only three squeezers, one SFG stage, and two beam-splitters, instead of $K$ times those numbers.

\begin{acknowledgements}
This research was supported by AFOSR Grant No.~FA9550-14-1-0052.
QZ thanks Aram Harrow for discussion of the Schur transform.
\end{acknowledgements}

\begin{center}{\Large\bf Supplemental Material}
\end{center}%

\section{Qubit approximation for sum-frequency generation}
\label{theory}
Our objective in this section is to develop the qubit approximation for the sum-frequency generation (SFG) process in which $M$ zero-mean, signal-idler mode pairs $\{(\hat{a}_{S_m},\hat{a}_{I_m}) : 1 \le m \le M\}$ and a single sum-frequency mode $\hat{b}$ undergo Schr\"{o}dinger evolution for $t\ge 0$ under the interaction Hamiltonian
\begin{equation}
\hat{H}_I = \hbar g\sum_{m=}^M (\hat{b}^\dagger\hat{a}_{S_m}\hat{a}_{I_m} + \hat{b}\hat{a}_{S_m}^\dagger\hat{a}_{I_m}^\dagger),
\label{Hinteract}
\end{equation}
where $M\gg 1$.
We will assume permutation-invariant initial conditions for the signal-idler mode pairs, so that the following $t\ge 0$ averages are all independent of the mode indices:
\begin{subequations}
\begin{align}
&n_s(t) \equiv \langle \hat{a}_{S_m}^\dagger\hat{a}_{S_m}\rangle_t, \,\,
n_i(t) \equiv \langle \hat{a}_{I_m}^\dagger\hat{a}_{I_m}\rangle_t \\ 
&C(t) \equiv \langle \hat{a}_{S_m}\hat{a}_{I_m}\rangle_t\\
&n_{si}(t) \equiv \langle \hat{a}_{S_m}^\dagger\hat{a}_{S_m}\hat{a}_{I_m}^\dagger\hat{a}_{I_m}\rangle_t\\
&G(t) \equiv \langle \hat{a}_{S_m}^\dagger\hat{a}_{S_n}\hat{a}_{I_m}^\dagger\hat{a}_{I_n}\rangle_t, \mbox{ for $m\neq n$}\\ 
&F(t) \equiv \langle \hat{a}_{S_m}^\dagger\hat{a}_{I_m}^\dagger\hat{b}\rangle.
\end{align}
\end{subequations}
In what follows we will use $b(t) = \langle \hat{b}\rangle_t$ and $n_b(t) = \langle \hat{b}^\dagger\hat{b}\rangle_t$ to denote the mean field and average photon number of the sum-frequency mode at time $t$, with $b(0) = 0$ and $n_b(0) = 0$ being their initial conditions.  

Let $\hat{\rho}(t)$ be the joint density operator for the $2M+1$ modes undergoing the Eq.~\eqref{Hinteract} interaction.  Now suppose that the total average photon number at $t=0$ is small, i.e., $M[n_s(0) + n_i(0)] + n_b(0) \ll 1$, as is the total phase-sensitive cross correlation, i.e., $M|C(0)|^2 \ll 1$. It follows that  $\hat{\rho}(t)$ has a number-state representation that, without appreciable loss of accuracy, can be truncated to the tensor product of the qubit Hilbert spaces spanned by the $2M+1$ modes' vacuum ($|0\rangle$) and single-photon ($|1\rangle$) states.  Within this qubit approximation, and using the special form of Eq.~\eqref{Hinteract}, we have that $\hat{\rho}(t)$'s only non-zero matrix elements, to lowest order in $n_s(t)$, $n_i(t)$ and $C(t)$, are 
\begin{subequations}
\ba
\langle \mathbf{0}|\hat{\rho}(t)|\mathbf{0}\rangle &=&1-O(n_i,n_s)\\
\langle \mathbf{10}_m|\hat{\rho}(t)|\mathbf{10}_m\rangle &=& n_s(t)\\
\langle \mathbf{01}_m|\hat{\rho}(t)|\mathbf{01}_m\rangle &=&n_i(t)\\
\langle \mathbf{11}_m|\hat{\rho}(t)|\mathbf{11}_m\rangle &=& n_{si}(t)\\
\langle \mathbf{11}_m|\hat{\rho}(t)|\mathbf{0}\rangle &=& C(t)\\
\langle \mathbf{11}_n|\hat{\rho}(t)|\mathbf{11}_m\rangle &=& G(t), \mbox{ for $m\neq n$}\\
\langle \mathbf{1}_b|\hat{\rho}(t)|\mathbf{1}_b\rangle &=& n_{b}(t)\\
\langle \mathbf{1}_b|\hat{\rho}(t)|\mathbf{0}\rangle&=& b(t)\\
\langle \mathbf{1}_b|\hat{\rho}(t)|\mathbf{11}_m\rangle &= & F(t),
\ea
\label{moments}
\end{subequations}
and the complex conjugates of the $C(t)$, $G(t)$, $b(t)$, and $F(t)$ terms.   
In Eqs.~\eqref{moments} we have introduced the compact notations 
\begin{subequations}
\begin{align}
|\mathbf{0}\rangle &= |0\rangle_b\bigotimes_{1\le m\le M}(|0\rangle_{S_m}|0\rangle_{I_m})\\
|\mathbf{jk}_m\rangle &= |0\rangle_b\bigotimes(|j\rangle_{S_m}|k\rangle_{I_m})\bigotimes_{\begin{array}{c}\scriptstyle 1\le n \le M\\ \scriptstyle n\neq m\end{array}}(|0\rangle_{S_n}|0\rangle_{I_n}\rangle)\\
|\mathbf{1}_b\rangle &= |1\rangle_b\bigotimes_{1\le m\le M}(|0\rangle_{S_m}|0\rangle_{I_m}),
\end{align}
\end{subequations}
with all the states on the right being number states whose subscripts identify the modes to which they apply.  
Note that, in general, trace operations are required to calculate the moments shown on the right in Eqs.~\eqref{moments}, but the qubit approximation obviates that need.

The full Hamiltonian for the SFG process is 
\be
\hat{H}=\hat{H}_0+\hat{H}_I,
\ee
where
\be
\hat{H}_0=\hbar \sum_{m=1}^{M}(\omega_{S_m} \hat{a}_{S_m}^\dagger \hat{a}_{S_m}+  \omega_{I_m} \hat{a}_{I_m}^\dagger \hat{a}_{I_m})+\hbar \omega_b \hat{b}^\dagger \hat{b},
\ee
and the mode-pair frequencies satisfy $\omega_{S_m}+\omega_{I_m}=\omega_b$. This Hamiltonian's matrix elements are easy to obtain in the qubit approximation, from which the Schr\"odinger equation, 
\be
\frac{{\rm d}\hat{\rho}(t)}{{\rm d}t}=\frac{1}{i\hbar}[\hat{H},\hat{\rho}(t)]
\ee
yields the following set of equations in the interaction picture:
\begin{subequations}
\ba
\frac{{\rm d}C(t)}{{\rm d}t}&=&-ig b(t)\\
\frac{{\rm d}b(t)}{{\rm d}t} &=&-igM C(t)\\
\frac{{\rm d}G(t)}{{\rm d}t} &=&-2g{\rm Im}[F(t)]\\
\frac{{\rm d}F(t)}{{\rm d}t}&=&ig[n_b(t)-n_{si}(t)-(M-1)G(t)]\\
\frac{{\rm d}n_{si}(t)}{{\rm d}t} &=&-2g{\rm Im}[F(t)]\\
\frac{{\rm d}n_b(t)}{{\rm d}t} &=&2Mg{\rm Im}[F(t)].
\ea
\end{subequations}

With the initial conditions given earlier, plus $G(0) = |C(0)|^2$, we obtain the following $t\ge 0$ solutions:
\begin{subequations}
\begin{align}
C(t)&=C(0)\cos (\sqrt{M}gt)
\label{cfinal0}
\\
b(t)&=-i \sqrt{M} C(0) \sin ( \sqrt{M}gt)
\label{bfinal0}
\\
n_b(t)&=[M|C(0)|^2+n_s(0)n_i(0)]\sin^2(\sqrt{M}gt)
\label{nbfinal0}
\\
n_{si}(t)&=(1-1/M)n_s(0)n_i(0) \nonumber \\
&\,\,+[|C(0)|^2+n_s(0)n_i(0)/M]\cos^2 (\sqrt{M}gt)\\
F(t)&=-i\sqrt{M}[|C(0)|^2+n_s(0)n_i(0)/M]\nonumber \\
&\,\, \times\sin (2\sqrt{M}gt)/2
\label{Ffinal0}\\
G(t)&=n_{si}(t)-n_i(0)n_s(0)\\
n_s(t)&=n_s(0)-n_b(t)/M, n_i(t)=n_i(0)-n_b(t)/M.
\label{nsnifinal0}
\end{align}
\end{subequations}
Equations~(3) from the paper are obtained from Eqs.~\eqref{cfinal0}--\eqref{nbfinal0}, and \eqref{nsnifinal0} by imposing the additional restriction $|C(0)|^2 \ll n_s(0)$ and $n_i(0)$. 

We see from Eqs.~\eqref{cfinal0} and \eqref{bfinal0} that any initial phase-sensitive cross correlation, $C(0)$, between the signal and idler modes will be completely converted to a sum-frequency-mode mean field, $b(t)$, at times $t_\ell=\ell\pi/2\sqrt{M}g$, for $\ell$ a non-negative odd integer.
Furthermore, Eq.~\eqref{Ffinal0} shows that $F(t_\ell) =0$.  Because $F(t)$ is the only qubit-approximation moment involving correlation between the signal-idler mode pairs and the sum-frequency mode, its vanishing at $t_\ell$ implies that the sum-frequency mode is not entangled with the signal-idler mode pairs at those times.  In addition, because $M \gg 1$, we have $n_{si}(t_\ell)\simeq n_s(0)n_i(0)$, and $G(t_\ell)\simeq 0$.  

At this point we know that duration-$t_\ell$ SFG eliminates any initial phase-sensitive cross correlation between the signal and idler in a mode pair, \emph{and} it neither entangles the signal-idler mode pairs with the sum-frequency mode, nor does it produce appreciable correlation between different signal-idler mode pairs.  As a result, we can approximate its effect on each signal-idler mode pair as a two-mode squeezing (TMS) operation characterized by the symplectic transformation $S[C(0)]$ \cite{footnote}.  Alternatively, in the Heisenberg picture with $\{\hat{a}_{S_m}(0),\hat{a}_{I_m}(0)\}$ and 
$\{\hat{a}_{S_m}(t_\ell),\hat{a}_{I_m}(t_\ell)\}$ being the initial and final mode-pair operators for duration-$t_\ell$ SFG, our TMS approximation is
\ba
\hat{a}_{S_m}(t_\ell)&=&\sqrt{1+|C(0)|^2}\hat{a}_{S_m}(0)-C(0) \hat{a}^\dagger_{I_m}(0)\nonumber\\
\hat{a}_{I_m}(t_\ell)&=&\sqrt{1+|C(0)|^2}\hat{a}_{I_m}(0)-C(0) \hat{a}^\dagger_{S_m}(0).
\label{Srexample}
\ea

\begin{figure*}
\centering
\includegraphics[width=1\textwidth]{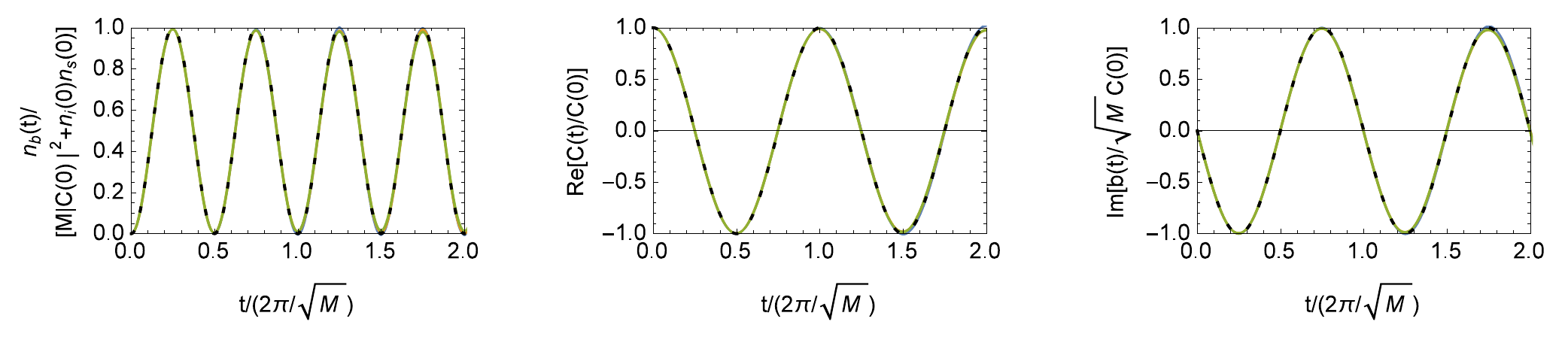}
\caption{Comparison between numerical solutions to the Schr\"odinger equation for $n_b(t)/[M|C(0)|^2 + n_s(0)n_i(0)]$, ${\rm Re}[C(t)/C(0)]$, and ${\rm Im}[b(t)/\sqrt{M}C(0)]$ (shown as curves) and the corresponding results obtained from Eqs.~\eqref{cfinal0}--\eqref{nbfinal0} (shown as points).   The assumed initial conditions were $n_i(0)=0.002$, $n_s(0)=0.0025$, and $C(0)=-0.0015$.  Calculations were performed for $M=1,2,3$ using $g=1$ and truncating the density operator's number-state expansion at 4 for $M=1,2,$ and 3 for $M=3$.  Our qubit approximation predicts the plotted moments should be independent of $M$, and our numerical solutions bear this out in that the curves for $M = 1,2,3$ are almost indistinguishable.}
\label{Fig1}
\end{figure*}

To justify our leading-order solutions, especially for $n_{si}(t)$, which is of the order of $n_i(0)n_s(0)$ and $|C(0)|^2$, we can include the next higher-order terms in $\hat{\rho}(t)$.  Owing to the form of our Hamiltonian, however, these terms do not contribute to the evolution of $n_{si}(t)$.  We have also compared the solutions in Eqs.~\eqref{cfinal0}--\eqref{nbfinal0} with numerical solutions of the Schr\"odinger equation for $M = 1,2, 3$ with $n_s(0), n_i(0) \ll 1$, $|C(0)|^2\ll n_s(0),n_i(0)$, and found excellent agreement, see, e.g., Fig.~\ref{Fig1}.

\section{Binary Mixed-State Discrimination}
Implementing a minimum error-probability receiver for deciding between two equally-probable possible states, $\hat{\rho}_0$ and $\hat{\rho}_1$, falls into the realms of quantum state discrimination \cite{Chefles_2000,Barnett_09,Barnett_1998,Andersson_2002,Anthony_1998,Chen_2015,BecerraF_2013,Becerra_2015,Mosley_2006,Clarke_2001,Tsujino_2011,Takeoka_2006,Takeoka_2005,Nair_2014}.  The minimum error-probability decision rule was given by Helstrom \cite{Helstrom1969} as:   
(1) perform the $u(\hat{\rho}_1-\hat{\rho}_0)$ measurement, where $u(\cdot)$ is the unit-step function; and (2) decide $\hat{\rho}_1$ if the measurement's outcome yields 1 and decide $\hat{\rho}_0$ otherwise.  Helstrom also showed that this decision rule's error probability was $P_H = (1-\sum_n \lambda^{(+)}_n)/2$, where the $\{\lambda_n^{(+)}\}$ are the non-negative eigenvalues of $\hat{\rho}_1-\hat{\rho}_0$.  

Unfortunately, Helstrom's minimum error-probability decision rule does not provide an explicit prescription for its experimental realization. However, when $\hat{\rho}_0$ and $\hat{\rho}_1$ are each comprised of $M$ independent, identically-distributed states, then running the Schur transform on a quantum computer will provide an efficient realization \cite{Bacon_2006} for achieving the quantum Chernoff bound (QCB) \cite{Audenaert2007,Pirandola2008} on $P_H$, which is known to be exponentially tight.  A simpler approach, consisting of local operations and classical communication (LOCC), suffices when the states in question are pure~\cite{Acin_2005}, but for the general mixed-state case it is known that collective operations are necessary to asymptotically achieve minimum error-probability performance \cite{LOCC_NO_GO,Bandyopadhyay_2011}.  Thus, for quantum-illumination (QI) target detection we must go beyond LOCC-based sub-optimum receivers \cite{Guha2009} in order to realize QI's full performance advantage over coherent-state target detection.  Our paper does so \emph{without} recourse to a full-up quantum computer.

\section{Binary Coherent-State Discrimination}
\label{Coherent_sec}
In this section, we consider the well-studied problem of minimum error-probability discrimination between equally-likely coherent states $|\alpha\rangle$ and $|-\alpha\rangle$, and use it to motivate our employing SFG to realize minimum error-probability QI target detection. Before going into the details, a brief historical review is worthwhile.  

In 1973, Kennedy described the first receiver for the preceding coherent-state discrimination problem whose error-probability exponent matched that of Helstrom's optimum decision rule~\cite{Kennedy_1972}. His analysis was couched in semiclassical terms, i.e., shot-noise theory, but its equivalent quantum description, given later in \cite{ShapiroPOVM}, uses displacement by $\alpha$ to null the mean field when the $|-\alpha\rangle$ state was present.   Photon counting and deciding $|\alpha\rangle$ was present if and only if a non-zero count is obtained then results in an error probability of $e^{-4|\alpha|^2}/2$, as compared to $P_H = [1-\sqrt{1-\exp(-4|\alpha|^2)}]/2 \approx e^{-4|\alpha|^2}/4$ for $|\alpha|^2 \gg 1$.  Later, Dolinar modified the Kennedy receiver to include a photodetection feedback scheme in which the pre-detection displacement operation becomes time-varying ~\cite{dolinar_processing_1973}.  Using shot-noise theory, he gave an explicit form for the feedback that enables his receiver's error probability to equal $P_H$, implying that, in quantum terms, it realizes the Helstrom measurement.  

More recently, other reception schemes have led to better performance than the Kennedy receiver. In the improved Kennedy receiver \cite{Gaussian_limit,improved_Kennedy_demo}, a fixed but optimized pre-detection displacement is employed, but the resulting error probability, while lower than $e^{-4|\alpha|^2}/2$, exceeds $P_H$. The Sasaki-Hirota receiver \cite{Sasaki_Hirota_receiver} achieves $P_H$ performance by replacing the Dolinar receiver's feedback with a single pre-detection unitary operation, but this single unitary is non-Gaussian, and it requires a nonlinearity of arbitrarily high order.  A more general result for arbitrary pure-state discrimination is also very enlightening \cite{Acin_2005}, for its nice Bayesian interpretation.

Now let us turn to the details for discriminating between equally likely $\hat{\rho}_1 = |\alpha\rangle\langle \alpha|$ and $\hat{\rho}_0 = |-\alpha\rangle\langle -\alpha|$.  
Defining $\hat{\rho}_1-\hat{\rho}_0$ to have the eigenket-eigenvalue decomposition $\sum_n(\lambda_n^{(+)}|\lambda_n^{(+)}\rangle \langle \lambda^{(+)}_n| + 
\lambda_n^{(-)}|\lambda_n^{(-)}\rangle \langle \lambda^{(-)}_n|)$ in terms of its positive and negative eigenvalues, the Helstrom measurement is equivalent to the positive operator-valued measurement (POVM) $\{\hat{\Pi}_1,\hat{I}-\hat{\Pi}_1\}$ with $\hat{\Pi}_1 = \sum_n |\lambda_n^{(+)}\rangle \langle \lambda^{(+)}_n|$ and $\hat{I}$ being the identity operator.

To gain valuable insight into the Kennedy receiver we start from the easily demonstrated result
\begin{equation}
\hat{D}(\alpha)(\hat{\rho}_1-\hat{\rho}_0)\hat{D}^\dagger(\alpha) = 
|2\alpha\rangle\langle 2\alpha| - |0\rangle\langle 0|,
\end{equation}
where $\hat{D}(\cdot)$ is the displacement operator.  Because the displacement operation is unitary, we can achieve $P_H$ performance \emph{after} performing that unitary if we take our POVM $\{\hat{\Pi}_1,\hat{I}-\hat{\Pi}_1\}$ to be the one for minimum error-probability discrimination between equally likely $\hat{D}(\alpha)\hat{\rho}_1\hat{D}^\dagger(\alpha)$ and $\hat{D}(\alpha)\hat{\rho}_0\hat{D}^\dagger(\alpha)$.  Using the basis $\{\ket{0^\bot}=(\ket{2\alpha}-e^{-2|\alpha|^2}\ket{0})/\sqrt{1-e^{-4|\alpha|^2}}, \ket{0}\}$, we then get
\begin{equation}
\label{intuition}
\hat{\Pi}_1 = \left(\begin{array}{cc}
(1-\sqrt{1-e^{-4|\alpha|^2}})/2 &  e^{-2|\alpha|^2}/2\\
 e^{-2|\alpha|^2}/2 & (1+\sqrt{1-e^{-4|\alpha|^2}})/2
\end{array} \right)
\end{equation}
for that optimum POVM.  
Equation~\eqref{intuition} shows that $\hat{\Pi}_1 \approx \hat{I}-|0\rangle \langle 0|$ to leading order when $|\alpha|^2 \gg 1$, which coincides with Kennedy receiver's POVM $\{\hat{I}-|0\rangle \langle 0|,|0\rangle\langle 0|\}$ \cite{ShapiroPOVM}.  This is the intuition for the Kennedy receiver:  forcing the vacuum state to be one of the hypothesized states so that a non-zero photon count enables error-free rejection of that state. 

The Sasaki-Hirota receiver extends the Kennedy-receiver paradigm by preceding a photon-number resolving (PNR) measurement with a unitary $\hat{U}$ chosen to make $|0\rangle\langle 0|$ one of the POVM elements for optimum discrimination between equally likely $\hat{U}\hat{\rho}_1\hat{U}^\dagger$ and $\hat{U}\hat{\rho}_0\hat{U}^\dagger$. This understanding motivates our use of SFG for the collective measurement in QI target detection:  SFG produces a vacuum (non-vacuum) sum-frequency output when low-brightness signal-idler pairs at its input have zero (non-zero) phase-sensitive cross correlation.  Moreover, because Dolinar needed to augment the Kennedy receiver with an optimum feedback law to achieve minimum error-probability operation, we should not be surprised that SFG will require the addition of a feed-forward structure to realize minimum error-probability QI target detection.
 
\section{QI target detection for weak signal-idler pairs}
As a preface to our treatment of the paper's QI target-detection problem, let us consider the simpler scenario in which the background light has low brightness but the problem is otherwise the same as in the paper.  In particular, for equally-likely target absence of presence ($h = 0$ or 1, respectively), we observe $M \gg 1$ returned-signal/retained-idler mode pairs, $\{\hat{c}_{S_m},\hat{c}_{I_m}\}$, that, conditioned on $h=j$, are in independent, identically-distributed, zero-mean, Gaussian states with Wigner covariance matrix
\begin{align}
& 
{\mathbf{\Lambda}}_j =
\frac{1}{4}
\left(
\begin{array}{cccc}
(2N_B+1) {\mathbf I}&2C_p{\mathbf Z}\delta_{1j}\\
2C_p {\mathbf Z}\delta_{1j}&(2N_S+1){\mathbf I}
\end{array} 
\right),
&
\label{CovReturnIdler_main}
\end{align}
where ${\mathbf I} = {\rm diag}(1,1)$, ${\mathbf Z}={\rm diag}(1,-1)$, $\delta_{ij}$ is the Kronecker delta function, $C_p = \sqrt{\kappa N_S(N_S+1)}$ is the phase-sensitive cross correlation that exists when the target is present, and $\kappa N_S\ll N_S \ll N_B \ll 1$.   Pirandola considered this problem in the $\kappa\sim 1$ quantum-reading context~\cite{Qreading}, but the quantum advantage over coherent-state operation that he found there vanishes when $\kappa \ll 1$.  So, our goal in this section will be to see how our SFG receiver can approach QCB performance, even though that performance will not be better than that achievable with coherent states.   Toward that end, we will content ourselves with asymptotic analytical results, reserving for later the full evaluation of the $N_B \gg 1$ scenario, in which QI target detection with a quantum-optimum receiver enjoys 6\,dB higher error-probability exponent than that of coherent-state operation.  

First let us exhibit the QCB for QI target detection in the weak signal, weak idler scenario under consideration here.  Using Pirandola and Lloyd's symplectic diagonalization technique~\cite{Pirandola2008}, it is easily shown that
\begin{equation}
P_{\rm QCB} \simeq \exp(-MC_p^2)/2\simeq \exp(-M\kappa N_S)/2,
\end{equation}
is that quantum Chernoff bound.  Next, we note that the optical parametric amplifier receiver~\cite{Guha2009} and the dual-homodyne receiver~\cite{Qreading} both achieve sub-optimum performance:
\begin{equation}
P_{\rm OPA} \simeq \exp(-MC_p^2/2)/2\simeq \exp(-M\kappa N_S/2)/2,
\end{equation}
and
\begin{equation}
P_{\rm Hom} 
\simeq \exp(-MC_p^2/2)/2\simeq \exp(-M\kappa N_S/2)/2,
\end{equation}
respectively, making their error-probability exponents 3\,dB inferior to the one for optimum QI reception.  Now let us show how SFG can recover that missing 3\,dB.

Because $1 \le  M|C_p|^2 \le 10$ at QCB values of interest, we cannot directly apply the qubit approximation for the SFG process to all $M$ mode pairs, even though the returned signal and retained idler both have low brightness.  Instead, we shall divide those $M$ mode pairs into $M/K$ subsets, with $1\ll K \ll M$ and $K|C_p|^2 \ll 1$, and have each subset undergo SFG.  When the target is absent, the $M/K$ sum-frequency outputs will all be in independent thermal states with average photon number $N_SN_B$.  When the target is present, those sum-frequency outputs will all be in coherent states with mean field $\sqrt{K}\,C_p$ that are embedded in independent thermal backgrounds with average photon number $N_SN_B$.  Coherent combining of the $M/K$ sum-frequency outputs then leads to the following binary state-discrimination task for distinguishing between target absence and presence.  Is the resulting state thermal with average photon number $N_SN_B$ (target absent), or is it a coherent state with mean field $\sqrt{M}\,C_p$ that is embedded in a thermal background with average photon number $N_SN_B$ (target present)?  The SFG receiver answers this question by counting photons in the coherently-combined sum-frequency output, and declaring target present if and only if one or more counts are obtained.  The resulting error probability is easily shown to be
\begin{equation}
P_{\rm SFG} \simeq [\exp(-MC_p^2) + N_SN_B]/2,
\end{equation}
which becomes
\begin{equation}
P_{\rm SFG} \simeq \exp(-MC_p^2) /2 \simeq \exp(-M\kappa N_S)/2,
\end{equation}
for $MC_p^2 \ll -\ln(N_SN_B)$.
By analogy with the Dolinar receiver, we expect that augmenting the weak-signal/weak-idler SFG receiver with an appropriate feed-forward structure could approach the Helstrom bound for this target-detection problem.

\begin{figure}
\centering
\includegraphics[width=0.5\textwidth]{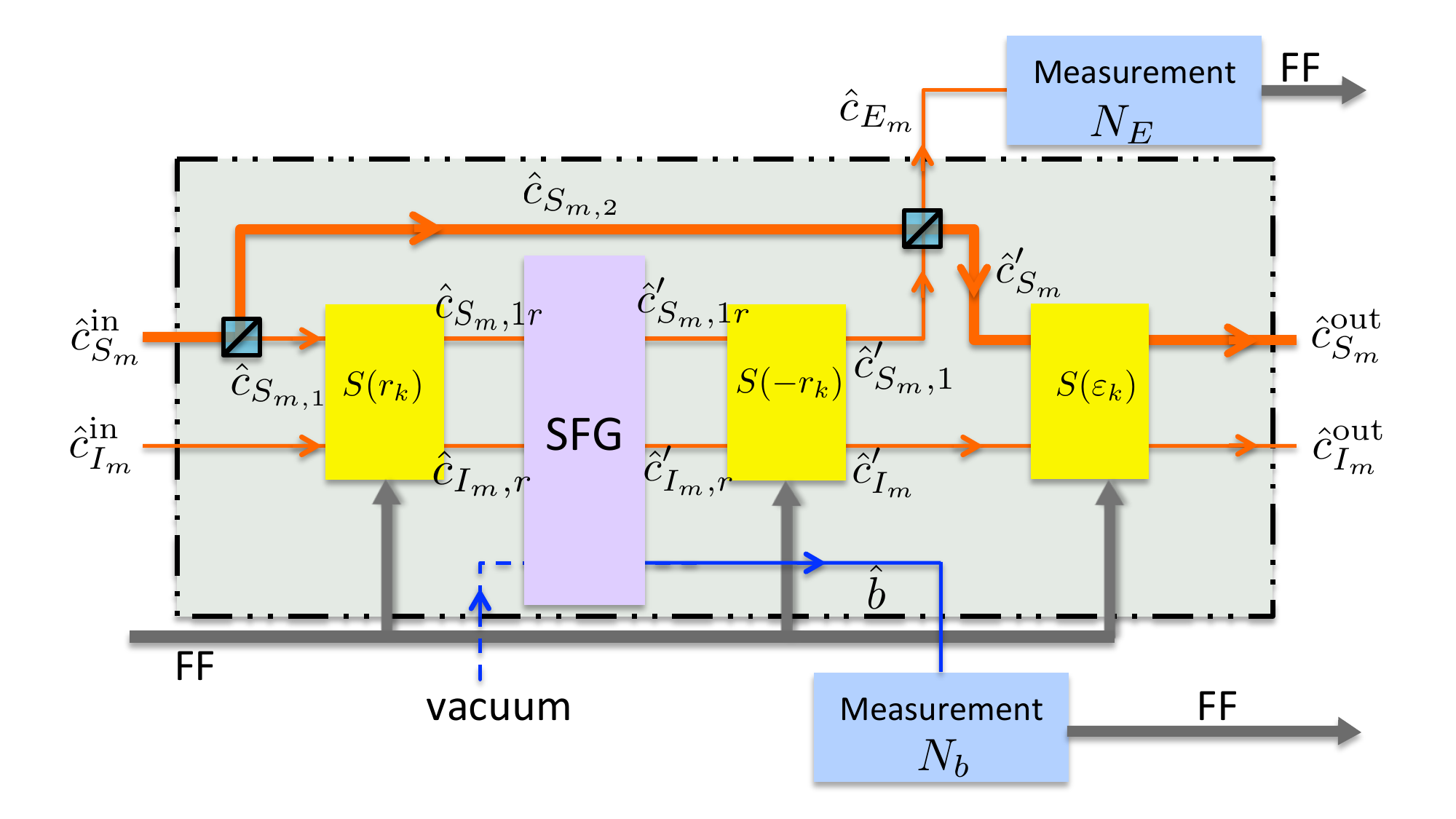}
\caption{Schematic of the FF-SFG receiver's $k$th cycle.  $S(r_k)$, $S(-r_k)$, and $S(\varepsilon_k)$: two-mode squeezers;  SFG:  sum frequency generation; FF:  feed-forward. 
}
\label{Fig2}
\end{figure}

\section{Single-Cycle Analysis of the FF-SFG Receiver}
In this section we return to the QI target-detection problem in its advantageous, $\kappa N_S \ll N_S \ll 1 \ll N_B$, setting.  As explained in the paper, we use signal slicing so that each SFG interaction in a $K$-cycle sequence transpires in the weak signal, weak idler regime.  As alluded to in the preceding section, we embellish these SFG interactions with feed-forward elements to enable the overall receiver to approach $P_H$ performance.  In order to evaluate the FF-SFG receiver's error probability, the immediate task is working through the functioning of the receiver's $k$th cycle.  The structure of that cycle, shown in the paper's Fig.~1, is reproduced here in Fig.~\ref{Fig2}.  In this figure we use ``in'' and ``out'' superscripts on the cycle's input and output signal and idler modes, in lieu of the ``$(k)$'' and ``$(k+1)$'' superscripts that appeared in the paper's Fig.~1.  We also suppress all other ``$(k)$'' superscripts, and we introduce some additional notation that will be needed in what follows.  

The inputs to the $k$th cycle are the $M\gg 1$ signal-idler mode pairs $\{(\hat{c}_{S_m}^{\rm in},\hat{c}_{I_m}^{\rm in})\}$ whose average photon numbers and phase-sensitive cross correlation are \cite{footnote2} $
n_s^{\rm in} \equiv \braket{\hat{c}_{S_m}^{{\rm in}\dagger}\hat{c}_{S_m}^{\rm in}} \gg 1,
n_i^{\rm in } \equiv \braket{\hat{c}_{I_m}^{{\rm in}\dagger}\hat{c}_{I_m}^{\rm in}} \ll 1,
C_{si}^{\rm in} \equiv \braket{\hat{c}_{S_m}^{\rm in}\hat{c}_{I_m}^{\rm in}}
$.  For $k=0$, the Wigner covariance matrix in the paper's Eq.~(1) gives us $n_s^{\rm in} = N_B \gg 1$, $n_i^{\rm in} = N_S \ll 1$, and $C_{si}^{\rm in}|_{h=1} = C_p = \sqrt{\kappa N_S(N_S + 1)} \ll 1$.

Because we need to exploit SFG in the regime wherein its qubit approximation is valid, we route the $\{\hat{c}_{S_m}^{\rm in}\}$ to a beam splitter with transmissivity $\eta \ll 1$ that slices off a low-brightness portion of this light, $\{\hat{c}_{S_m,1}\}$, given by
\begin{equation}
\hat{c}_{S_m,1}=\sqrt{\eta}\,\hat{c}_{S_m}^{\rm in}+\sqrt{1-\eta}\,\hat{c}_{v_m},
\end{equation}
where the $\{\hat{c}_{v_m}\}$ are vacuum-state modes.  That beam splitter's other outputs, $\{\hat{c}_{S_m,2}\}$, given by
\begin{equation}
\hat{c}_{S_m,2}=\sqrt{1-\eta}\,\hat{c}_{S_m}^{\rm in}-\sqrt{\eta}\,\hat{c}_{v_m},
\label{bs1}
\end{equation}
will later be merged with the SFG-processed signal modes from the $k$th cycle to provide the input signal modes for the $(k+1)$th cycle.

The $\{\hat{c}_{S_m,1}\}$ and the $\{\hat{c}_{S_m,2}\}$ have average photon numbers $\eta n_s^{\rm in}$ and $(1-\eta)n_s^{\rm in}$,  phase-insensitive cross correlation $\langle \hat{c}_{S_m,1}^\dagger\hat{c}_{S_m,2}\rangle = \sqrt{\eta(1-\eta)}\,n_s^{\rm in}$, and their phase-sensitive cross correlations with the $\{\hat{c}_{I_m}^{\rm in}\}$ modes are $\langle \hat{c}_{S_m,1}\hat{c}_{I_m}^{\rm in}\rangle = \sqrt{\eta}\,C_{si}^{\rm in}$ and $\langle \hat{c}_{S_m,2}\hat{c}_{I_m}^{\rm in}\rangle = \sqrt{1-\eta}\,C_{si}^{\rm in}$.
 
Following this signal slicing, the $\{(\hat{c}_{S_m,1},\hat{c}_{I_m}^{\rm in})\}$ mode pairs undergo the TMS operation $S(r_k)$, with $r_k$ given by Eq.~(5) from the paper, yielding outputs
\begin{align}
\hat{c}_{S_m,1r} &=\sqrt{1+r_k^2}\,\hat{c}_{S_m,1}-r_k\hat{c}_{I_m}^{{\rm in}\dagger} \\
\hat{c}_{I_m,r}&=\sqrt{1+r_k^2}\,\hat{c}_{I_m}^{\rm in}-r_k\hat{c}_{S_m,1}^\dagger,
\end{align}
which can be rewritten as
\begin{align}
\hat{c}_{S_m,1r}& =\sqrt{\eta}(\hat{c}_{S_m}^{\rm in}-f(k)\hat{c}_{I_m}^{{\rm in}\dagger})+  {\rm h.o.t.} \\
\hat{c}_{I_m,r}&=\hat{c}_{I_m}^{\rm in}-f(k)\sqrt{\eta}\hat{c}_{v_m}+{\rm h.o.t.},
\end{align}
where h.o.t.~denotes terms that are higher order in $\eta$, and $f(k) \equiv r_k/\sqrt{\eta}$ is weakly-dependent on $\eta$.  Throughout the analysis that follows, \emph{all} h.o.t.~terms will be included, but, except when we use big-$O$ notation to indicate the scale of terms that are not shown, we will only report the leading-order terms.  That said, we have the following moment results to leading order in $\eta$:
\begin{align}
&\braket{\hat{c}_{S_m,1r}^\dagger\hat{c}_{S_m,1r}}=\eta n_s^{\rm in}+\eta[f^2(k)-2f(k) C_{si}^{\rm in}]+{\rm h.o.t.}&
\nonumber\\ &\braket{\hat{c}_{I_m,r}^\dagger \hat{c}_{I_m,r}}=n_i^{\rm in}+\eta[f^2(k)-2f(k) C_{si}^{\rm in}]+{\rm h.o.t.}&
\nonumber\\ &\braket{\hat{c}_{S_m,1r}\hat{c}_{I_m,r}}=\sqrt{\eta}\,[C_{si}^{\rm in}-f(k)]+{\rm h.o.t.}&
\nonumber\\
&\braket{\hat{c}_{S_m,1r}^\dagger \hat{c}_{I_m,r}}=0&
\nonumber\\
&\braket{\hat{c}_{S_m,1r}\hat{c}_{S_m,2}^\dagger}=\sqrt{(1-\eta)\eta}[n_s^{\rm in}-f(k) C_{si}^{\rm in}]+{\rm h.o.t.}&
\nonumber\\
&\braket{\hat{c}_{I_m,r}\hat{c}_{S_m,2}}=\sqrt{1-\eta}[C_{si}^{\rm in}-f(k)\eta n_s^{\rm in}]+{\rm h.o.t.}.&
\end{align}
where the h.o.t.~are of order $O(\eta^{3/2})$.
For notational compactness, it is convenient to employ $r\equiv \sqrt{\eta}[C_{si}^{\rm in}-f(k)] = \sqrt{\eta}\,C_{si}^{\rm in}-r_k$, with the understanding that $r$ has implicit dependence on $k$.   

Next, the $\{(\hat{c}_{S_m,1r},\hat{c}_{I_m,r})\}$ mode pairs undergo duration-$t_\ell$ SFG. Because all these modes have low brightness, and we can choose $\eta$ small enough that, despite having $M\gg 1$, their time evolution will be governed by Eqs.~\eqref{cfinal0}--\eqref{nbfinal0} with $t=t_\ell$.  Hence, any non-zero $\braket{\hat{c}_{S_m,1r}\hat{c}_{I_m,r}}$ will be annihilated by the SFG process through conversion to the mean field of sum-frequency output.  
Furthermore, as shown in Sec.~\ref{theory}, that sum-frequency output will not be entangled with the output signal-idler mode pairs. So, as we did in Sec.~I, we can make a Gaussian approximation to the $t_\ell$-duration SFG, representing its signal-idler outputs, $\{(\hat{c}_{S_m,1r}',\hat{c}_{I_m,r}')\}$, as being the result of the TMS operation $S(r)$, viz.,
\ba
\hat{c}_{S_m,1r}^\prime&=&\sqrt{1+r^2}\,\hat{c}_{S_m,1r}-r\hat{c}_{I_m,r}^{\dagger}
\nonumber\\
&=&\sqrt{\eta}(\hat{c}_{S_m}^{\rm in}-C_{si}^{\rm in}\hat{c}_{I_m}^{{\rm in}\dagger})+ {\rm h.o.t.}\nonumber\\
\hat{c}_{I_m,r}^\prime&=&\sqrt{1+r^2}\hat{c}_{I_m,r}-r\hat{c}_{S_m,1r}^\dagger\nonumber\\
&=&\hat{c}_{I_m}^{\rm in}-C_{si}^{\rm in}\sqrt{\eta}\,\hat{c}_{v}^\dagger+ {\rm h.o.t.}.
\ea

The sum-frequency mode, $\hat{b}$, emerging from the SFG process will be in a coherent state $|\sqrt{M}r\rangle$ that is immersed in an $M$-independent thermal background of average photon number $\eta n_s^{\rm in}n_i^{\rm in}$ that will turn out to satisfy $\eta n_s^{\rm in}n_i^{\rm in}\simeq \eta N_BN_S \ll Mr^2$ \cite{footnote2}. We can also obtain the following quantities to leading order in $\eta$.
\ba
\braket{\hat{c}_{S_m,1r}^\prime\hat{c}_{I_m,r}^\prime}&=&0
\nonumber\\
\braket{\hat{c}_{S_m,1r}^{\prime\dagger}\hat{c}_{S_m,1r}^\prime}&=&\eta n_s^{\rm in}-\eta |C_{si}^{\rm in}|^2+ {\rm h.o.t.}
\nonumber\\
\braket{\hat{c}_{I_m,r}^{\prime\dagger}\hat{c}_{I_m,r}^\prime}&=&n_{i}^{\rm in}-\eta |C_{si}^{\rm in}|^2+ {\rm h.o.t.}
\nonumber\\
\braket{\hat{c}_{S_m,1r}^\prime \hat{c}_{S_m,2}^\dagger}&=& \sqrt{\eta(1-\eta)}(n_s^{\rm in}-|C_{si}^{\rm in}|^2)+ {\rm h.o.t.}
\nonumber\\
\braket{\hat{c}_{I_m,r}^\prime\hat{c}_{S_m,2}}&=&\sqrt{1-\eta}\,C_{si}^{\rm in}(1-\eta n_s^{\rm in})+ {\rm h.o.t.}
\ea

The signal-idler outputs from the SFG process are used as inputs to another TMS operation,  $S(-r_k)$.  Its purpose is to make the total average photon number in the $\{\hat{c}_{E_m}\}$ signal modes that will be measured in the $k$th cycle about the same as that of the sum-frequency mode, $\hat{b}$.  In particular, we have that
\ba
\hat{c}_{S_m,1}^{\prime}&=&\sqrt{1+r_k^2}\,\hat{c}_{S_m,1r}^\prime+r_k\hat{c}_{I_m,r}^{\prime\dagger}
\nonumber\\
&=&\sqrt{\eta}(\hat{c}_{S_m}^{\rm in}+[f(k)-C_{si}^{\rm in}]\hat{c}_{I_m}^{{\rm in}\dagger})+ {\rm h.o.t.}
\nonumber\\
\hat{c}_{I_m}^\prime&=&\sqrt{1+r_k^2}\hat{c}_{I_m,r}^\prime+r_k\hat{c}_{S_m,1r}^{\prime\dagger}
\nonumber\\
&=&\hat{c}_{I_m}^{\rm in}+[f(k)-C_{si}^{\rm in}]\sqrt{\eta}\,\hat{c}_{v}^\dagger+ {\rm h.o.t.},
\ea
from which we can obtain the following quantities to leading order in $\eta$:
\begin{align}
&\braket{\hat{c}_{S_m,1}^{\prime\dagger}\hat{c}_{S_m,1}^{\prime}}={\eta}n_s^{\rm in}-\eta[{|C_{si}^{\rm in}|}^2-f^2(k)]+ {\rm h.o.t.}&
\nonumber\\
&\braket{\hat{c}_{I_m}^{\prime\dagger}\hat{c}_{I_m}^\prime}=n_i^{\rm in}-\eta[{|C_{si}^{\rm in}|}^2-f^2(k)]+ {\rm h.o.t.}&
\nonumber\\
&\braket{\hat{c}_{S_m,1}^{\prime}\hat{c}_{I_m}^\prime}=f(k)\sqrt{\eta}+ {\rm h.o.t.}&
\nonumber\\
&\braket{\hat{c}_{S_m,1}^{\prime}\hat{c}_{S_m,2}^{\dagger}}=\sqrt{\eta(1-\eta)}[n_s^{\rm in}-{|C_{si}^{\rm in}|}^2+f(k) C_{si}^{\rm in}]+ {\rm h.o.t.}&
\nonumber\\
&\braket{\hat{c}_{S_m,2}\hat{c}_{I_m}^{\prime}}=\sqrt{1-\eta}[C_{si}^{\rm in}+{\eta}n_s^{\rm in}(f(k)-C_{si}^{\rm in})]+ {\rm h.o.t.}&
\end{align}

After the $S(-r_k)$ operation, we interferometrically combine the $\{\hat{c}_{S_m,1}^{\prime}\}$ modes with the $\{\hat{c}_{S_m,2}\}$ modes on a beam splitter with transmissivity $\eta$, and obtain outputs
\ba
\hat{c}_{S_m}^\prime&=&\sqrt{\eta}\,\hat{c}_{S_m,1}^{\prime}+\sqrt{1-\eta}\,\hat{c}_{S_m,2}
\nonumber\\
&=&\hat{c}_{S_m}^{\rm in}+\eta[f(k)-C_{si}^{\rm in}]\hat{c}_{I_m}^{{\rm in}\dagger}+ {\rm h.o.t.}
\nonumber\\
\hat{c}_{E_m}&=&\sqrt{1-\eta}\,\hat{c}_{S_m,1}^{\prime}-\sqrt{\eta}\,\hat{c}_{S_m,2}
\nonumber\\
&=&\sqrt{\eta}\,[f(k)-C_{si}^{\rm in}]\hat{c}_{I_m}^{{\rm in}\dagger}+ {\rm h.o.t.},
\label{CEeqn}
\ea
which implies the following moments to leading order in $\eta$:
\begin{align}
&n_{E_m}\equiv \braket{\hat{c}_{E_m}^{\dagger}\hat{c}_{E_m}}=r^2+O(\eta^2,\eta n_i^{\rm in} r)&
\nonumber\\
&\braket{\hat{c}_{S_m}^{\prime\dagger}\hat{c}_{S_m}^\prime}=n_s^{\rm in}-2\eta[{|C_{si}^{\rm in}|}^2-f(k) C_{si}^{\rm in}]+ {\rm h.o.t.}&
\nonumber\\
&\braket{\hat{c}_{S_m}^\prime \hat{c}_{I_m}^\prime}= C_{si}^{\rm in}[1-\eta(1+n_s^{\rm in})]+f(k) \eta (1+n_s^{\rm in})+ {\rm h.o.t.}&
\label{extra_term}
\end{align}
Here we see that the total average photon number, $\sum_{m=1}^M\langle \hat{c}_{E_m}^\dagger\hat{c}_{E_m}\rangle$, in the signal-mode states that will be measured in the $k$th cycle do indeed have about the same $Mr^2$ average photon number as the sum-frequency mode $\hat{b}$.  
One can also check that the $\{\hat{c}_{E_m}\}$ modes are in independent, identically-distributed, zero-mean states, whose Wigner covariance matrix is that of a thermal state. 

Equation~\eqref{extra_term} shows the presence of an undesired extra term, $f(k) \eta (1+n_s^{\rm in})$, in the phase-sensitive cross correlation between the $\{\hat{c}_{S_m}'\}$ and $\{\hat{c}_{I_m}'\}$ modes.  To eliminate this term, we employ yet another TMS operation, $S(\varepsilon_k)$, with $\varepsilon_k=\eta f(k) \ll1$ on the strong signal and weak idler: 
\begin{align}
\hat{c}_{S_m}^{\rm out}&=\sqrt{1+\varepsilon_k^2}\,\hat{c}_{S_m}^\prime-\varepsilon_k \hat{c}_{I_m}^{\prime\dagger} \nonumber \\
&=\hat{c}_{S_m}^{\rm in}-\eta C_{si}^{\rm in}\hat{c}_{I_m}^{{\rm in}\dagger}+ {\rm h.o.t.}  
\nonumber\\
\hat{c}_{I_m}^{\rm out}&= \sqrt{1+\varepsilon_k^2}\,\hat{c}_{I_m}^\prime-\varepsilon_k \hat{c}_{S_m}^{\prime\dagger}
\nonumber\\
&=\hat{c}_{I_m}^{\rm in}+\sqrt{\eta}(f(k)-C_{si}^{\rm in})\hat{c}_{v}^\dagger-\eta C_{si}^{\rm in} \hat{c}_{S_m}^{{\rm in}\dagger}+ {\rm h.o.t.}
\label{modeevo}
\end{align}
It is now straightforward to obtain
\begin{subequations}
\begin{align}
n_s^{\rm out}&\equiv \braket{\hat{c}_{S_m}^{{\rm out}\dagger} \hat{c}_{S_m}^{\rm out}}
\nonumber\\
&=n_s^{\rm in}-2\eta {|C_{si}^{\rm in}|}^2+O(\eta^2,{\eta}n_i^{\rm in}{|C_{si}^{\rm in}|}^2)\\
n_i^{\rm out}&\equiv \braket{\hat{c}_{I_m}^{{\rm out}\dagger} \hat{c}_{I_m}^{\rm out}}  \nonumber \\
& =n_i^{\rm in}-\eta ({|C_{si}^{\rm in}|}^2-f^2(k)+2f(k)  C_{si}^{\rm in}) \nonumber \\
&+ O(\eta^2,{\eta}n_i^{\rm in}{|C_{si}^{\rm in}|}^2)\\
C_{si}^{\rm out}&\equiv \braket{\hat{c}_{S_m}^{\rm out)} \hat{c}_{I_m}^{\rm out}}
=C_{si}^{\rm in}[1-\eta(1+n_s^{\rm in})]
\nonumber\\
&+ O(\eta^2,\eta n_s^{\rm in} n_i^{\rm in}C_{si}^{\rm in},\eta |C_{si}^{\rm in}|^3)
\end{align}
\end{subequations}

For the $k=0$ cycle we have $n_s^{\rm in} \equiv n_s^{(0)} \gg1$, $n_i^{\rm in} \equiv n_i^{(0)} \ll 1$ and $|C_{si}^{\rm in}|^2 \equiv |C_{si}^{(0)}|^2 \ll n_i^{(0)}$.  Using these initial conditions plus $\eta\ll1$, we find that
\begin{subequations}
\ba
n_s^{\rm out}&\simeq& n_s^{\rm in}\\
n_i^{\rm out}&\simeq& n_i^{\rm in}\\
C_{si}^{\rm out}&\simeq &C_{si}^{\rm in}[1-\eta(1+n_s^{\rm in})].
\label{ckeqn}
\ea
\end{subequations}
Replacing the ``in'' and ``out'' superscripts with ``($k$)`` and ``($k+1$)'' in Eqs.~\eqref{ckeqn}, and using the preceding initial conditions, gives the paper's Eqs.~(4).    

At this juncture, we have all the results we need to determine the measurement statistics that the paper uses in the FF-SFG receiver's update equation, i.e., in its Eq.~(6).  What we need there is $P_{BE}(N_b^{(k-1)},N_E^{(k-1)};j,r^{(k-1)}_{\tilde{h}_{k-1}})$ for $1\le k \le K$ and $j=0,1$, where $r^{(k-1)}_{\tilde{h}_{k-1}}$ is given by the paper's Eq.~(5), $\tilde{h}_{k-1}$ is the minimum error-probability decision as to which hypothesis is true based on observations prior to cycle $k-1$, $N_b^{(k-1)}$ is the photon count in the $\hat{b}^{(k-1)}$ mode, and $N_E^{(k-1)}$ is the total photon count in the $\{\hat{c}^{(k-1)}_{E_m}\,\}$ modes.  Because the sum-frequency mode $\hat{b}^{(k-1)}$ is generated by SFG in the qubit-approximation regime, the results of Sec.~I imply that
\begin{eqnarray}
\lefteqn{P_{BE}(N_b^{(k-1)},N_E^{(k-1)};j,r^{(k-1)}_{\tilde{h}_{k-1}}) = } \nonumber \\
&& P_B(N_b^{(k-1)};j,r^{(k-1)}_{\tilde{h}_{k-1}})P_E(N_E^{(k-1)};j,r^{(k-1)}_{\tilde{h}_{k-1}}).
\end{eqnarray}

To obtain $P_E(N_E^{(k-1)};j,r^{(k-1)}_{\tilde{h}_{k-1}})$, we use Eq.~\eqref{ckeqn} and conclude that the $\{\hat{c}_{E_m}^{(k-1)}\}$ are a collection of independent, identically-distributed modes in thermal states with average photon number $r^2$, whence
\begin{align}
P_E(N_E^{(k-1)};j,r^{(k-1)}_{\tilde{h}_{k-1}}) &= \left(\begin{array}{c}N_E^{(k-1)}+M-1\\ M-1\end{array}\right) \nonumber \\
& \times \frac{\tilde{r}^{2N_E^{(k-1)}}}{(1+\tilde{r}^2)^{N_E^{(k-1)} + M}},
\end{align}
where $\tilde{r}\equiv r^{(k-1)}_{\tilde{h}_{k-1}\oplus j}$ with $\oplus$ denoting exclusive or.  
Because $\tilde{r}^2 \ll 1$, we will have $N_E^{(k-1)} \ll M$, which makes $N_E^{(k-1)}$'s distribution approximately binomial, viz.,
\begin{align}
P_E(N_E^{(k-1)};j,r^{(k-1)}_{\tilde{h}_{k-1}}) &\simeq \left(\begin{array}{c}M\\ N_E^{(k-1)}\end{array}\right) \nonumber\\ 
&\times \tilde{r}^{2N_E^{(k-1)}}(1+\tilde{r}^2)^{M-N_E^{(k-1)}}.
\end{align}

To obtain $P_B(N_b^{(k-1)};j,r^{(k-1)}_{\tilde{h}_{k-1}})$ we use the fact that the sum-frequency mode $\hat{b}^{(k)}$ in the coherent state $|\sqrt{M}\tilde{r}\rangle$ immersed in a thermal background with average photon number $\eta n_sn_i=\eta N_BN_S$.  Because this state is classical \cite{Shapiro2009a}, we can use semiclassical (shot-noise) theory \cite{Gagliardi} to show that $N_b^{(k-1)}$ is Laguerre distributed:
\begin{eqnarray}
\lefteqn{\hspace*{-.1in}P_B(N_b^{(k-1)};j,r^{(k-1)}_{\tilde{h}_{k-1}}) = \exp\!\left(-\frac{M\tilde{r}^2}{\eta N_BN_S + 1}\right)}\nonumber \\
&\hspace*{-.15in}\times& \frac{(\eta N_BN_S)^{N_E^{(k-1)}}}{(1+\eta N_BN_S)^{N_E^{(k-1)}+1}}L_{N_E^{(k-1)}}\!\left(-\frac{M\tilde{r}^2}{\eta N_BN_S + 1}\right)\!,
\end{eqnarray}
where $L_p(\cdot)$ is the $p$th Laguerre polynomial.

\section{Intuition behind the feed-forward update rule}
In this section, we provide an intuitive explanation for the feed-forward Bayesian update rule given in the paper's Eqs.~(5) and (6).  Because the distinction between equally-likely target absence (hypothesis $h=0$) and target presence (hypothesis $h=1$) in QI lies in whether ($h=1$) or not ($h=0$) there is a phase-sensitive cross correlation between the returned and retained light, it might seem that optimum quantum reception for this problem should use SFG to convert the phase-sensitive cross correlation that signifies target presence into a maximum-strength coherent-state component at the sum frequency that would not be there were the target absent.  Neglecting, for simplicity, the sum-frequency output's weak thermal-state component that is present under both hypotheses makes this cross-correlation-nulling receiver analogous to the Kennedy receiver for equally-likely, binary phase-shift-keyed coherent states.  In particular, both saturate their hypothesis test's quantum Chernoff bound, but neither achieves their test's Helstrom bound.  

It is easy, for the Kennedy receiver, to see why at weak signal levels lower error probability can be realized with a displacement operation that does \emph{not} perform nulling.  Consider coherent-state binary phase-shift keying (BPSK) in which:  (1) the received field's coherent state is equally likely to be $|(-1)^{h+1}\alpha\rangle$ for bit values $h=0$ or 1; (2) a mean-field displacement by $\beta$ is performed followed by ideal photon counting; and (3) $\tilde{h}=0$ is declared as the received bit value if no counts are obtained and $\tilde{h}=1$ is declared otherwise.  This improved Kennedy receiver has an error probability given by~\cite{Gaussian_limit,improved_Kennedy_demo}
\be
P_E=\frac{1}{2}\left[1-e^{-\left(\alpha-\beta\right)^2}+e^{-\left(\alpha+\beta\right)^2}\right],
\ee
whose minimum, for $|\alpha|\ll 1$, occurs at $\beta = 1/\sqrt{2} \gg |\alpha|$, whereas for $|\alpha| \gg 1$ that minimum occurs very close to the $\beta = \alpha$ value employed by the Kennedy receiver.  Even with its optimum displacement, however, this improved Kennedy receiver fails to achieve the Helstrom bound for coherent-state BPSK.  The Dolinar receiver gets to that limit by observing the photon-count record over time and evolving its displacement according to a Bayesian-update rule obtained from dynamic programming.  This rule produces a smoothly-varying displacement versus time except for discrete jumps that occur whenever a count is recorded.  As the posterior probabilities for $h=0$ and 1 evolve, they become increasingly asymmetric, and the Dolinar receiver's displacement gets closer to nulling the mean field for one of the two possible states.  Let us see how these considerations play out in our FF-SFG receiver, where the nulling in question is for the phase-sensitive cross correlation.  

\begin{figure*}
\centering
\subfigure[]{
\includegraphics[width=0.4\textwidth]{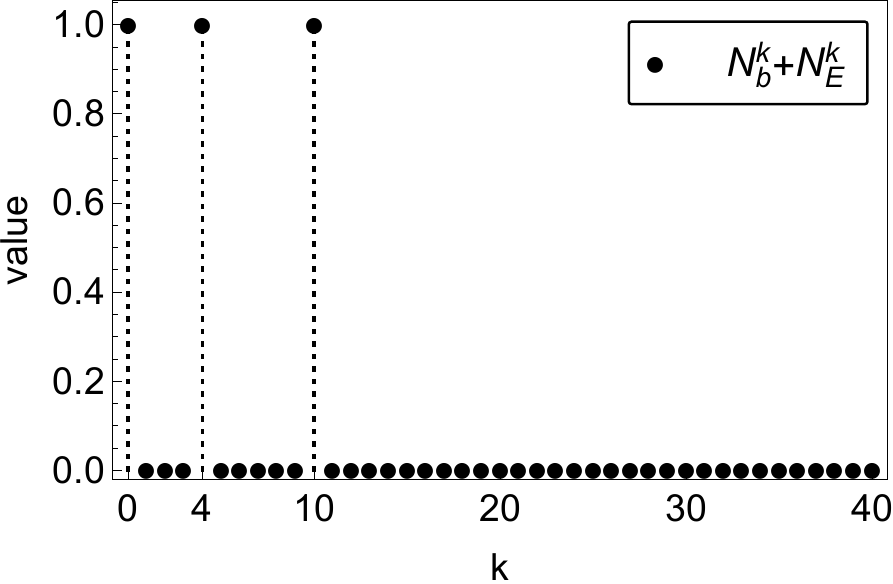}
\label{3a}
}
\subfigure[]{
\includegraphics[width=0.4\textwidth]{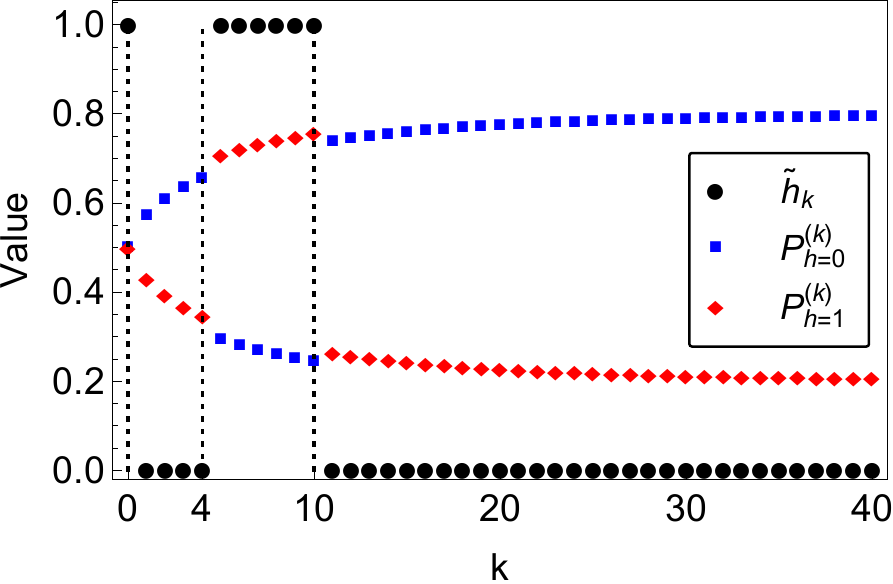}
\label{3b}
}
\subfigure[]{
\includegraphics[width=0.4\textwidth]{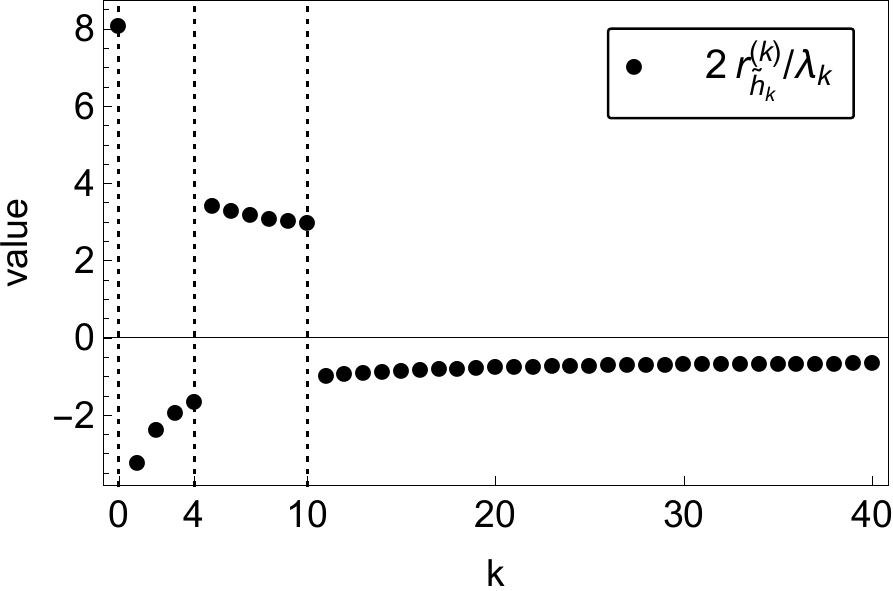}
\label{3c}
}
\subfigure[]{
\includegraphics[width=0.4\textwidth]{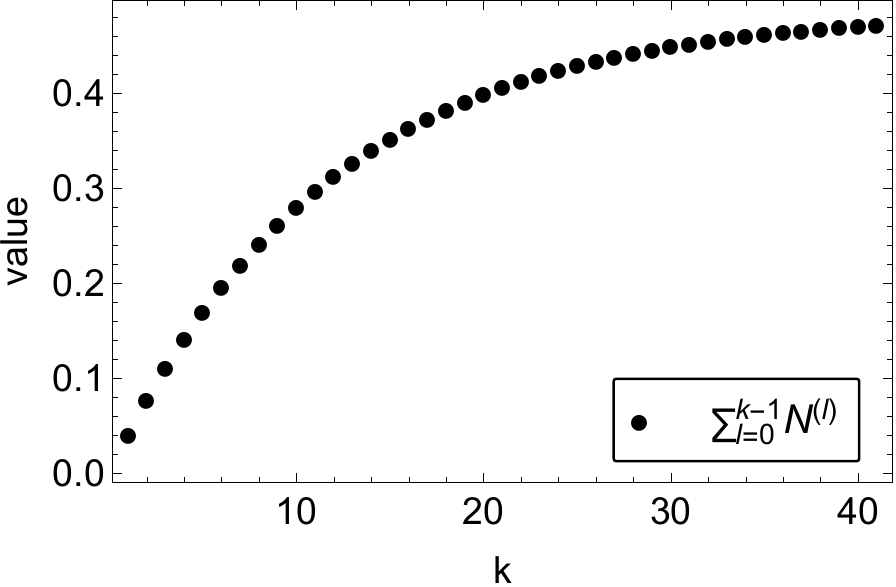}
\label{3d}
}
\caption{Single simulation run of the feed-forward update rule for target absence with $M=10^7$ and other parameters as in the paper's Fig.~2(a). All plots are versus the cycle number $k$.  (a) The photon count $N_b^{(k)}+N_E^{(k)}$ in cycle $k$. (b) The prior probabilities $\{P^{(k)}_{h=j} : j = 0, 1\}$ and the tentative decision $\tilde{h}_k$ for cycle $k$. (c) The scaled squeezing parameter $2r^{(k)}_{\tilde{h}_k}/\lambda_k$, given by the paper's Eq.~(5), for cycle $k$. (d) The coherent part of the total mean photon number, $\sum_{\ell=0}^{k-1}N^{(\ell)}$.
}
\label{Fig3}
\end{figure*} 

Returning to the paper's Eq.~(5), we see that $r^{(k)}_{\tilde{h}_k}$ contains the factor
\begin{equation}
\sigma_k \equiv \left\{1-\exp\!\left[-2 M\!\left(\sum_{\ell=0}^{k}\lambda_\ell^2-\lambda_k^2/2\right)\right]\right\}^{-1/2},
\end{equation}
where $\lambda_k^2 = \eta C_p^2 [1-(1-\eta(1+N_B))]^{2k}$ is monotonically decreasing with increasing $k$ because $\eta N_B \ll 1$.  For $M \gg 1$, it follows that $\sigma_k$ monotonically decreases with increasing $k$, starting from a large value at $k=0$, and approaching unity as $k\rightarrow \infty$.  Consequently, the FF-SFG receiver never uses its two-mode squeezing operations to completely eliminate the phase-sensitive cross correlation, even when the tentative decision $\tilde{h}_k$ in the $k$th cycle is correct. Deriving the paper's Eq.~(5) requires dynamic programming, as done for the Dolinar receiver~\cite{Assalini2011}, but the monotonic decrease in $\sigma_k$ towards unity can be understood as follows. For the initial cycles of FF-SFG reception, i.e., when $\sigma_k \gg 1$, the prior probabilities are nearly equal, so we have that  $r^{(k)}_{\tilde{h}_k} \approx (-1)^{\tilde{h}_k}\lambda_k\sigma_k/2$.   This squeezing value does \emph{not} eliminate the phase-sensitive cross correlation between the returned and retained light that is presumed to be present at the outset of the $k$th cycle.  As cycles proceed, however, the prior probabilities become increasingly asymmetric.  We then get $\sigma_k$ approaching unity, so that $r^{(k)}_{\tilde{h}_k} \approx \lambda_k[1-(-1)^{\tilde{h}_k}]$, which \emph{is} attempting to null that phase-sensitive cross correlation.  Then, when the thermal-state contributions to $N_b^{(k)}$ and $N_E^{(k)}$ are neglected, detection of one photon unambiguously indicates that the present decision $\tilde{h}_k$ must be wrong. 
The optimum way to use the $\sigma_k>1$ that we have for small $k$ values requires the Bayesian update rule from the paper's Eq.~(6).  This equation can be understood as follows.

The number of detected photons $N_b^{(k-1)},N_E^{(k-1)}$ and the two-mode squeezing $r_{\tilde{h}_{k-1}}^{(k-1)}$ applied in the $(k-1)$th cycle are used to derive the $(k-1)$th cycle's {\em posterior} probabilities, which will serve as the $k$th cycle's \emph{prior} probabilities. The terms entering into this update procedure have the following interpretations:
\begin{itemize}
\item $P_{h=j}^{(k-1)}$ is the prior probability of hypothesis $j$ being true given the measurements from cycles 0 through $k-2$.

\item $P_{BE}(N_b^{(k-1)},N_E^{(k-1)};j,r_{\tilde{h}_{k-1}}^{(k-1)})$ is the conditional joint probability of observing $N_b^{(k-1)}$ photons in the $\hat{b}^{(k-1)}$ mode and a total of $N_E^{(k-1)}$ photons in all the $\{\hat{c}_{E_m}^{(k-1)}\}$ modes conditioned on hypothesis $j$ being true and that $r_{\tilde{h}_{k-1}}^{(k-1)}$ of two-mode squeezing was applied in the $(k-1)$th cycle.

\item $P_{h=j}^{(k)}$ is the prior probability of hypothesis $j$ given the measurements from cycles 0 through $k-1$. It equals  the posterior probability of the hypothesis $j$ being true given the number of photons observed and the amount of two-mode squeezing applied in the $(k-1)$th cycle.
\end{itemize}
At the outset of the $k$th cycle, the tentative decision $\tilde{h}_{k}$ is taken to be the $j$ value that maximizes $P_{h=j}^{(k)}$. 

To illustrate the FF-SFG receiver's decision process we ran a simulation for target absence ($h=0$) using the same parameters as the paper's Fig.~2(a) with $M=10^7$.  That $M$ value gives relatively high error probability ($>0.1$), which affords us a case in which multiple decision ($\tilde{h}_k$) reversals occur with increasing $k$.  The results of our simulation are shown in Fig.~\ref{Fig3}.  

At $k=0$, we used a random guess to get $\tilde{h}_0 =1$, as shown in Fig.~\ref{3b}. This guess yielded a large positive $r^{(0)}_{\tilde{h}_0}$ value, i.e., a high-gain two-mode squeezing operation is performed, immediately giving rise to a photon-detection event, as shown in Fig.~\ref{3a}, that flipped the tentative decision to $\tilde{h}_1=0$ (see Fig.~\ref{3b}). Furthermore, $r^{(1)}_{\tilde{h}_1}$ became negative (see Fig.~\ref{3c}). 

No photons were detected in the next three cycles ($k=1,2,3$), so that the prior probabilities $\{P_{h=0}^{(k)},P_{h=1}^{(k)}\}$  evolved smoothly through the Bayesian update, as seen in Fig.~\ref{3b}. Because $P_{h=0}^{(k)} > P_{h=1}^{(k)}$, we had $\tilde{h}_{k}=0$ in cycles 2, 3, and 4. 

A second photon was detected in cycle 4 (see Fig.~\ref{3a}), dramatically altering the prior probabilities for cycle 5. The updated prior probabilities then flip the decision to $\tilde{h}_{5}=1$.  The decision $\tilde{h}_{k}=1$ remains in force until a third photon was detected at $k=10$. This induces another dramatic correction to the prior probabilities that leads to $\tilde{h}_{11}=0$. Because no additional photons are detected after cycle 11, the prior probabilities continue to be smoothly updated until the termination condition is met. The final decision $\tilde{h}_{K}=0$ is correct in this simulation run.

\section{Choosing the Number of Cycles} 
Here we detail the procedure for choosing the number of cycles, $K$, used by the FF-SFG receiver.  In the paper we defined  
$\lambda_k\equiv \sqrt{\eta}\,C_p[1-\eta(1+N_B)]^k$ and $N_T^{(K)} \equiv 2M\sum_{k=0}^{K-1}\lambda_k^2$.  Then, for some pre-chosen $0< \epsilon \ll 1$, we took $K$ to be large enough that $N_T^{(K)}/N_T^{(\infty)} = 1-\epsilon$.  To see how this is accomplished, we proceed as follows. First, under the assumption that $h=1$, we rewrite Eq.~\eqref{ckeqn} as
\begin{equation}
C_{si}^{(k+1)}-C_{si}^{(k)}=-\lambda_k^2(1+N_B)/C_{si}^{(k)}.
\label{Clambda}
\end{equation}
Then, using this result, we get
\ba
N_T^{(K)}&=&-[2M/(1+N_B)]\sum_{k=0}^{K-1}(C_{si}^{(k+1)}-C_{si}^{(k)}) C_{si}^{(k)}
\nonumber
\\
&\simeq &
-\frac{2M}{(1+N_B)}\int_{C_p}^{C_{si}^{(K)}} {\rm d}C_{si}\,C_{si}
\nonumber
\\
&=&M[C_p^2-(C_{si}^{(K)})^2]/(1+N_B),
\ea 
and, because the cross correlation will be depleted in the limit $K\rightarrow \infty$, 
\be
N_T^{(\infty)}=MC_p^2/(1+N_B)=M\kappa N_S(1+N_S)/(1+N_B). 
\ee
This result reduces to $N_T^{(\infty)}\simeq M\kappa N_S/N_B$, as QI target detection has $N_S \ll 1$ and $N_B \gg 1$, and we get the termination expression given in the paper:  $N_T^{(K)} \simeq (1-\epsilon)M\kappa N_S/N_B$, because $2M\sum_{k=0}^{K-1}\lambda_k^2$ is the coherent contribution to $\sum_{k=0}^{K-1}\langle \hat{b}^{(k)\dagger}\hat{b}^{(k)}\rangle$.  It follows that the residual cross-correlation at termination is $C_{si}^{(K)}=\sqrt{\epsilon}\,C_p$.

Another use of Eq.~\eqref{Clambda} now allows us to obtain an explicit result for $K$.  In particular, we have that the incoherent contribution to $\sum_{k=0}^{K-1}\langle \hat{b}^{(k)\dagger}\hat{b}^{(k)}\rangle$ is
\begin{align}
K\eta N_BN_S &= -\sum_{k=0}^{K-1}(C_{si}^{(k+1)}-C_{si}^{(k)})/C_{si}^{(k)}
\nonumber
\\
&\simeq -\int_{C_p}^{\sqrt{\epsilon}\,C_p} \frac{{\rm d}C_{si}}{C_{si}}
 =-N_S\ln(\epsilon)/2,
\label{sumeq}
\end{align}
which can be rearranged to yield $K = -\ln(\epsilon)/2\eta N_B$, as stated in the paper, where $\eta \ll 1$ must be small enough to ensure that SFG's qubit approximation is valid. 

\section{Monte Carlo Simulations}
In this last section we describe how we performed the Monte Carlo simulations whose results were presented in the paper's Fig.~2.  To perform these simulations, we need the \emph{joint} statistics of the $\{\,(N_b^{(k)},N_E^{(k)})\}$ for all $k$.  In Sec.~V we only developed the \emph{marginal} (single-cycle) statistics for those photon counts, namely $\{P_{BE}(N_b^{(k)},N_E^{(k)};r^{(k)}_{\tilde{h}_{k}\oplus j})\}$, so there is some work left to do.  The independence of the sum-frequency mode $\hat{b}^{(k)}$ from the $\{c_{E_m}^{(k)}\}$ modes simplifies the task before us.  Furthermore, because each successive sum-frequency mode is generated from SFG using a different slice of the signal, it is appropriate to assume that all the $\{N_b^{(k)}\}$ are statistically independent, in which case their joint distribution is just the product of the marginals we derived in Sec.~V.  For the $\{\hat{c}_{E_m}^{(k)}\}$ modes there is statistical independence across different $m$ values but \emph{not} across different $k$ values.  Thus their joint photon-counting statistics require careful evaluation.   

To obtain the joint statistics of $\{N_E^{(k)} : 0\le k \le K-1\}$, we must recognize that there is very strong correlation across the $K$ cycles.  We proceed, therefore, by using Eq.~\eqref{CEeqn} with the inclusion of a vacuum-state mode ($\hat{v}^{(k)}_{E_m}$) that had been part of the ``h.o.t.,'' to obtain
\begin{equation}
\hat{c}_{E_m}^{(k)} \simeq \frac{\sqrt{\eta}\,C_{si}^{(k)}-r_k}{\sqrt{\eta}\,C_{si}^{(0)} - r_0}\,\hat{c}_{E_m}^{(0)} + \sqrt{1-\left|\frac{\sqrt{\eta}\,C_{si}^{(k)}-r_k}{\sqrt{\eta}\,C_{si}^{(0)} - r_0}\right|^2}\,\hat{v}^{(k)}_{E_m}.
\end{equation}
This beam-splitter-like relation plus the classical-state nature of thermal and vacuum states then allow us to generate $\{N_E^{(k)} : 0\le k \le K-1\}$ with the correct joint statistics by the following procedure.  

First, we recognize that if $\{\mu_m : 1 \le m \le M\}$ are independent, identically-distributed, exponential random variables with mean values $\langle \mu_m\rangle = |\sqrt{\eta}\,C_{si}^{(0)} -r_0|^2$, then, given the $\mu_{\rm tot} \equiv \sum_{m=1}^M\mu_m$, the $\{N_E^{(k)}\}$ are independent, Poisson-distributed random variables with mean values 
\begin{equation}\label{PoissonMeans}
E\!\left[N_E^{(k)} \mid \mu_{\rm tot} \right] = \left|\frac{\sqrt{\eta}\,C_{si}^{(k)}-r_k}{\sqrt{\eta}\,C_{si}^{(0)}-r_0}\right|^2\mu_{\rm tot}.
\end{equation}
Furthermore, because $M \gg 1$, we can take $\mu_{\rm tot}$ to be a Gaussian random variable with mean $M|\sqrt{\eta}\,C_{si}^{(0)}-r_0|^2$ and variance $M|\sqrt{\eta}\,C_{si}^{(0)}-r_0|^4$.  So, to obtain $\{N_E^{(k)} : 0\le k \le K-1\}$ with the correct joint statistics, we first generate a Gaussian random variable with that mean and variance, and then generate the $\{N_E^{(k)} : 0 \le k \le K-1\}$ as independent Poisson random variables whose mean values are given by Eq.~\eqref{PoissonMeans}.


\begin{thebibliography}{99}
\bibitem{footnoteDecision}For $h$ equally likely to be 0 or 1, we take $\tilde{h}_0$ to be 0 or 1 with equal probability.
\bibitem{footnote2}We will see, below, that $n_s^{\rm out} \simeq n_s^{\rm in}$ and $n_i^{\rm out} \simeq n_i^{\rm in}$, which justifies our asserting the $n_s^{\rm in} \gg 1$ and $n_i^{\rm in} \ll 1$ for the $k$th cycle, because those conditions hold for $k =0$.
\end{thebibliography}
\end{document}